\documentclass[twocolumn,epjc]{svjour3}
\pdfoutput=1
\usepackage{lineno}
\usepackage{authblk}
\usepackage[section]{placeins}
\usepackage{subfigure}
\usepackage{tabulary}
\usepackage{booktabs}
\usepackage{mathrsfs}
\usepackage{rotating}
\usepackage{float} 
\usepackage{url}
\usepackage{subfigure}
\usepackage{graphicx}
\usepackage{epstopdf}
\usepackage{rotating}
\usepackage{multirow}
\usepackage{xspace}
\usepackage{hyperref}

\smartqed  
\RequirePackage{graphicx}
\RequirePackage{amssymb}
\RequirePackage{placeins}
\RequirePackage{amsmath}

\graphicspath{ {./Plots/} }

\def\TeV{\ifmmode {\mathrm{\ Te\kern -0.1em V}}\else
                   \textrm{Te\kern -0.1em V}\fi}%
\def\GeV{\ifmmode {\mathrm{\ Ge\kern -0.1em V}}\else
                   \textrm{Ge\kern -0.1em V}\fi}%
\def\MeV{\ifmmode {\mathrm{\ Me\kern -0.1em V}}\else
                   \textrm{Me\kern -0.1em V}\fi}%
\def\keV{\ifmmode {\mathrm{\ ke\kern -0.1em V}}\else
                   \textrm{ke\kern -0.1em V}\fi}%
\def\eV{\ifmmode  {\mathrm{\ e\kern -0.1em V}}\else
                   \textrm{e\kern -0.1em V}\fi}%
\let\tev=\TeV
\let\gev=\GeV

\def\ifb{\mbox{fb$^{-1}$} \xspace}

\newcommand{\Wboson}{\ensuremath{W}}

\newcommand{\ZprimeM}{\ensuremath{M_{\Zprime}}}

\newcommand{\ttbar}{\ensuremath{t\overline{t}} }
\newcommand{\Zprime}{\ensuremath{Z'} }
\newcommand{\pt}{\ensuremath{p_T} }
\newcommand{\hpp}{\textsc{Herwig}\texttt{++} }
\newcommand{\eq}[1]{Eq.~\eqref{eq:#1}}

\newcommand{\subsec}[1]{Sec.~\ref{subsec:#1}}

\newcommand{\pythia}  {{\sc Pyth\-ia}}
\newcommand{\pythiasix}  {{\sc Pyth\-ia~6}}
\newcommand{\pythiaeight}  {{\sc Pyth\-ia~8}}

\newcommand{\LHC}{LHC}

\newcommand{\pp}{proton-proton}
\newcommand{\pu}{pile-up}
\newcommand{\PU}{Pile-up}
\newcommand{\pujet}{pile-up jet}
\newcommand{\qcdjet}{QCD jet}
\newcommand{\stojet}{stochastic jet}

\newcommand{\sqrts}[1]{\ensuremath{\sqrt{s} = #1} \TeV}
\newcommand{\Rpt}{\ensuremath{R_{p_T}}}
\newcommand{\ptmatch}{\ensuremath{p_T^{match}}}
\newcommand{\ptcorr}{\ensuremath{p_T^{corr}}}

\newcommand{\MB}{MB}
\newcommand{\pT}{\ensuremath{p_{\mathrm{T}}}}
\newcommand{\pTmin}{\ensuremath{p_{\mathrm{T}}^{\mathrm{min}}}}

\newcommand{\Npart}{\ensuremath{N_{\mathrm{part}}}}

\newcommand{\Ncoll}{\ensuremath{N_{\mathrm{coll}}}}
\newcommand{\axing}{\ensuremath{\langle\mu\rangle}}

\newcommand{\bjet}{\ensuremath{b}\mbox{-}{\rm jet}}
\newcommand{\FJ}{{\sc Fast\-Jet}}
\newcommand{\JVF}{\ensuremath{\mathcal{F}_{\mathrm{jvf}}}}

\newcommand{\kT}{\ensuremath{k_{\mathrm{T}}}}
\newcommand{\antikt}{{\rm anti-}\kT}

\newcommand{\as}{\alpha_s} 

\journalname{Eur. Phys. J. C}
\title{\vskip -1cm Boosted objects and jet substructure at the LHC}
\subtitle{Report of BOOST2012, held at IFIC Valencia, 23$^{rd}$-27$^{th}$ of July 2012.}
\titlerunning{Boosted objects at the LHC}
\authorrunning{BOOST2012 participants} 
\author[1]{\mbox{A. Altheimer}}
\author[2]{\mbox{A. Arce}} 
\author[3]{\mbox{L. Asquith}} 
\author[4]{\mbox{J. Backus Mayes}} 
\author[5]{\mbox{E. Bergeaas Kuutmann}} 
\author[6]{\mbox{J. Berger}} 
\author[2]{\mbox{D. Bjergaard}} 
\author[7]{\mbox{L. Bryngemark}} 
\author[8]{\mbox{A. Buckley}} 
\author[9]{\mbox{J. Butterworth}}
\author[10]{\mbox{M. Cacciari}} 
\author[9]{\mbox{M. Campanelli}} 
\author[11]{\mbox{T. Carli}} 
\author[12]{\mbox{M. Chala}}
\author[13]{\mbox{B. Chapleau}} \author[14]{\mbox{C. Chen}} 
\author[15]{\mbox{J.P. Chou}} 
\author[16]{\mbox{Th. Cornelissen}} 
\author[17]{\mbox{D. Curtin}} 
\author[18]{\mbox{M. Dasgupta}} 
\author[9]{\mbox{A. Davison}} 
\author[19]{\mbox{F. de Almeida Dias}} 
\author[20]{\mbox{A. de Cosa}} 
\author[11]{\mbox{A. de Roeck}} 
\author[8]{\mbox{C. Debenedetti}} 
\author[21]{\mbox{C. Doglioni}} 
\author[22]{\mbox{S.~D. Ellis}} 
\author[23]{\mbox{F. Fassi}} 
\author[24]{\mbox{J. Ferrando}} 
\author[16]{\mbox{S. Fleischmann}} 
\author[25]{\mbox{M. Freytsis}} 
\author[26]{\mbox{M.L. Gonzalez Silva}} 
\author[23]{\mbox{S. Gonzalez de la Hoz}} 
\author[21]{\mbox{F. Guescini}} 
\author[27]{\mbox{Z. Han}} 
\author[4]{\mbox{A. Hook}} 
\author[22]{\mbox{A. Hornig}} 
\author[4]{\mbox{E. Izaguirre}} 
\author[4]{\mbox{M. Jankowiak}} 
\author[28]{\mbox{J. Juknevich}} 
\author[23]{\mbox{M. Kaci}} 
\author[24]{\mbox{D. Kar}} 
\author[29]{\mbox{G. Kasieczka}} 
\author[30]{\mbox{R. Kogler}} 
\author[4]{\mbox{A. Larkoski}} 
\author[31]{\mbox{P. Loch}} 
\author[27]{\mbox{D. Lopez Mateos}} 
\author[32]{\mbox{S. Marzani}} 
\author[33]{\mbox{L. Masetti}} 
\author[34]{\mbox{V. Mateu}} 
\author[35]{\mbox{D.~W. Miller}} 
\author[36]{\mbox{K. Mishra}} 
\author[4]{\mbox{P. Nef}}
\author[24]{\mbox{K. Nordstrom}}
\author[23]{\mbox{E. Oliver Garcia}} 
\author[37]{\mbox{J. Penwell}} 
\author[38]{\mbox{J. Pilot}} 
\author[29]{\mbox{T. Plehn}} 
\author[39]{\mbox{S. Rappoccio}} 
\author[40]{\mbox{A. Rizzi}} 
\author[23]{\mbox{G. Rodrigo}} 
\author[41]{\mbox{A. Safonov}} 
\author[10,11]{\mbox{G.~P. Salam}} 
\author[23]{\mbox{J. Salt}} 
\author[29]{\mbox{S. Schaetzel}} 
\author[42]{\mbox{M. Schioppa}} 
\author[29]{\mbox{A. Schmidt}} 
\author[22]{\mbox{J. Scholtz}} 
\author[4]{\mbox{A. Schwartzman}} 
\author[27]{\mbox{M.~D. Schwartz}} 
\author[43]{\mbox{M. Segala}} 
\author[44]{\mbox{M. Son}}
\author[45]{\mbox{G. Soyez}} 
\author[32]{\mbox{M. Spannowsky}} 
\author[34]{\mbox{I. Stewart}} 
\author[46]{\mbox{D. Strom}}
\author[4]{\mbox{M. Swiatlowski}} 
\author[23]{\mbox{V. Sanchez Martinez}} 
\author[29]{\mbox{M. Takeuchi}} 
\author[34]{\mbox{J. Thaler}} 
\author[1]{\mbox{E. Thompson}} 
\author[36]{\mbox{N.~V. Tran}} 
\author[25]{\mbox{C. Vermilion}} 
\author[23]{\mbox{M. Villaplana}} 
\author[23]{\mbox{M. Vos}} 
\author[4]{\mbox{J. Wacker}} 
\author[47]{\mbox{J. Walsh}}

\affil[1]{Columbia University, Nevis Laboratory, Irvington, NY 10533, USA}
\affil[2]{Duke University, Durham, NC 27708, USA}
\affil[3]{Argonne National Laboratory, Lemont, IL 60439, USA}
\affil[4]{SLAC National Accelerator Laboratory, Menlo Park, CA 94025, USA}
\affil[5]{Deutsches Elektronen-Synchrotron, DESY, D-15738 Zeuthen, Germany}
\affil[6]{Cornell University, Ithaca, NY 14853, USA}
\affil[7]{Lund University, Lund, SE 22100, Sweden}
\affil[8]{University of Edinburgh, EH9 3JZ, UK}
\affil[9]{University College London, WC1E 6BT, UK}
\affil[10]{LPTHE, UPMC Univ.~Paris 6 and CNRS UMR 7589, Paris, France}
\affil[11]{CERN, CH-1211 Geneva 23, Switzerland}
\affil[12]{CAFPE and U. of Granada, Granada, E-18071, Spain}
\affil[13]{McGill University, Montreal, Quebec H3A 2T8, Canada}
\affil[14]{Iowa State University, Ames, Iowa 50011, USA}
\affil[15]{Rutgers University, Piscataway, NJ 08854, USA}
\affil[16]{Bergische Universitaet Wuppertal, Wuppertal, D-42097, Germany}
\affil[17]{YITP, Stony Brook University, Stony Brook, NY 11794-3840, USA}
\affil[18]{University of Manchester, Manchester, M13 9PL, UK}
\affil[19]{UNESP - Universidade Estadual Paulista, Sao Paulo, 01140-070, Brazil}
\affil[20]{INFN and University of Naples, IT80216, Italy}
\affil[21]{University of Geneva, CH-1211 Geneva 4, Switzerland}
\affil[22]{University of Washington, Seattle, WA 98195, USA}
\affil[23]{Instituto de F\'isica Corpuscular, IFIC/CSIC-UVEG, E-46071 Valencia, Spain}
\affil[24]{University of Glasgow, Glasgow, G12 8QQ, UK}
\affil[25]{Berkeley National Laboratory, University of California, Berkeley, CA 94720, USA}
\affil[26]{Universidad de Buenos Aires, AR-1428, Argentina}   
\affil[27]{Harvard University, Cambridge, MA 02138, USA}
\affil[28]{Weizmann Institute, 76100 Rehovot, Israel}
\affil[29]{Universitaet Hamburg, DE-22761, Germany}
\affil[30]{Universitaet Heidelberg, DE-69117, Germany}
\affil[31]{University of Arizona, Tucson, AZ 85719, USA}
\affil[32]{IPPP, University of Durham, Durham, DH1 3LE, UK}
\affil[33]{Universitaet Mainz, DE 55099, Germany}
\affil[34]{MIT, Cambridge, MA 02139, USA}
\affil[35]{University of Chicago, IL 60637, USA}
\affil[36]{Fermi National Accelerator Laboratory, Batavia, IL 60510, USA}
\affil[37]{Indiana University, Bloomington, IN 47405, USA}
\affil[38]{University of California, Davis, CA 95616, USA}
\affil[39]{Johns Hopkins University, Baltimore, MD 21218, USA}
\affil[40]{INFN and University of Pisa, Pisa, IT-56127, Italy}
\affil[41]{Texas A \& M University, College Station, TX 77843, USA}
\affil[42]{INFN and University of Calabria, Rende, IT-87036, Italy}
\affil[43]{Brown University, Richmond, RI 02912, USA}
\affil[44]{Yale University, New Haven, CT 06511, USA}
\affil[45]{CEA Saclay, Gif-sur-Yvette, FR-91191, France}
\affil[46]{University of Illinois, Chicago, IL 60607, USA}
\affil[47]{University of California, Berkeley, CA 94720, USA}
\begin{document}
\renewcommand\Affilfont{\textnormal\itshape\it\small}
\maketitle

\begin{abstract}
This report of the BOOST2012 workshop presents the results of four working groups that studied key aspects of jet substructure. We discuss the potential of the description of jet substructure in first-principle QCD calculations and study the accuracy of state-of-the-art Monte Carlo tools. Experimental limitations of the ability to resolve substructure are evaluated, with a focus on the impact of additional proton proton collisions on jet substructure performance in future LHC operating scenarios. A final section summarizes the lessons learnt during the deployment of substructure analyses in searches for new physics in the production of boosted top quarks. 
\end{abstract}

\section{Introduction}
\label{sec:intro}

With a centre-of-mass energy of 7~\tev{} in 2010 and 2011 and 
of 8~\tev{} in 2012 the LHC has pushed the energy frontier
well into the~\tev{} regime. Another leap in energy is expected 
with the start of the second phase of operation in 2014, when the 
centre-of-mass energy is to be increased to 13-14~\tev{}.
For the first time experiments produce large samples of $W$ and $Z$ 
bosons and top quarks with a transverse momentum $p_T$ that considerably 
exceeds their rest mass $m$ ($p_T \gg m$). The same is true also for the Higgs 
boson and, possibly, for as yet unknown particles with masses near
the electroweak scale.  
In this new kinematic regime, well-known particles are observed in unfamiliar 
ways. {\it Classical} reconstruction algorithms that rely on a one-to-one
jet-to-parton assignment are often inadequate,
in particular for hadronic decays of such boosted objects.

A suite of techniques has been developed to fully exploit the opportunities 
offered by boosted objects at the LHC. Jets are reconstructed 
with a much larger radius parameter to capture the energy of the complete 
(hadronic) decay in a single jet. The internal structure of these 
{\it fat} jets is a key signature to identify boosted objects among the 
abundant jet production at the LHC. Many searches use a 
variety of recently proposed substructure observables.
Jet grooming techniques\footnote{We refer to three related techniques as jet
grooming: filtering~~\cite{Butterworth:2008iy}, 
trimming~\cite{Krohn:2009th} and 
pruning~\cite{Ellis:2009me}. Unless stated otherwise, all studies
in this paper of these techniques apply a common set of parameters that
is widely used in the community.} 
improve the resolution of jet substructure 
measurements, help to reject background, 
and increase the resilience to the impact of 
multiple proton-proton interactions.

In July 2012 IFIC Valencia organized the 2012 edition~\cite{boost12} 
of the BOOST series of workshops, the main forum for the physics of 
boosted objects and jet substructure\footnote{Previous BOOST workshops
took place at SLAC (2009, ~\cite{boost09}), 
Oxford University (2010,~\cite{boost10}) and Princeton 
University University (2011,~\cite{boost11}). BOOST2013~\cite{boost13} 
was organized by the University of Arizona from August 12$^{th}$ to 16$^{th}$.}.
Working groups formed during the 2010 and 2011 workshops prepared 
reports~\cite{Abdesselam:2010pt,Altheimer:2012mn} that provide an
overview of the state of the field and an entry point to the 
now quite extensive literature and present new material prepared 
by participants. 
In this paper we present the report of the working groups set up 
during BOOST2012. Each contribution addresses an important aspect 
of jet substructure as a tool for the study of boosted objects at the LHC.

A good understanding of jet substructure is a prerequisite to further
progress. Predictions of jet substructure based on first-principle,
analytical calculations may provide a more precise description
of jet substructure and allow deeper insight. However, resummation 
of the leading logarithms in this case is notoriously difficult and the
predictions may be subject to considerable uncertainties.
In fact, one might ask:
\begin{itemize}
\item Can jet substructure be predicted by first-principle QCD calculations
and compared to data in a meaningful way? 
\end{itemize}  
The findings of the working group that was set up to evaluate 
the limitations and potential of the most popular approaches are 
presented in Section~\ref{sec:qcd}.

While progress toward analytical predictions continues, searches for 
boosted objects that employ jet substructure rely on 
the predictions of mainstream Monte Carlo models. It is therefore
vital to answer this question:
\begin{itemize}
\item How accurately is jet substructure described by state-of-the-art Monte Carlo tools?
\end{itemize}
The BOOST2010 report~\cite{Abdesselam:2010pt} provided a 
partial answer, based on pre-LHC tunes of several popular leading-order 
generators. After the valuable experience gained in the first three years of 
operation of the LHC, it seems appropriate to revisit this question
 in Section~\ref{sec:mc}. 

A further potential limitation to the performance of jet substructure 
is the level to which the detector response can be understood and modelled. 
Again, the first years of LHC operation have provided valuable experience 
on how well different techniques work in a realistic experimental environment. 
In particular, the impact of multiple proton-proton 
interactions (pile-up) on substructure measurement has been evaluated
exhaustively and mitigation schemes have been developed. 
Anticipating a sharp increase in the pile-up activity in future operating
scenarios of the LHC, one might worry that in the future the detector 
performance might be degraded considerably for the sensitive 
substructure analyses. A third working group was therefore given the 
following charge:
\begin{itemize}
\item  How does the impact of additional proton proton collisions 
limit jet substructure performance at the LHC, now and in future 
operating scenarios?  
\end{itemize}
Section~\ref{sec:pileup1} presents the contributions 
regarding jet reconstruction performance under 
extreme contributions, with up to 200 additional proton-proton collisions
in each bunch crossing. We present the prospects for fake jet
rates and the impact of pile-up on jet mass measurements 
under these conditions.

In the first years of operation of the LHC several groups in ATLAS and CMS
have deployed techniques specifically developed for the study of 
boosted objects in several analyses. Jet substructure has become 
an important tool in many searches for evidence for new physics. 
In Section~\ref{sec:top} we present the lessons learnt in several
studies of boosted top quark production that have been the first
to apply these techniques and answer the following question: 
\begin{itemize}
\item How powerful is jet substructure in studies of boosted top production, and how can it be made even more powerful?
\end{itemize}

We hope that the answers to the above questions prepared by the working
groups may shed some light on this rapidly evolving field.


\section{Measurements and first-principle QCD predictions for jet substructure}
\label{sec:qcd}




{\it Section prepared by the Working Group: 'Predictions and measurements of jet substructure observables', A. Davison, \underline{A. Hornig}, \underline{S. Marzani}, \underline{D.W. Miller}, G. Salam, M. Schwartz, I. Stewart, J. Thaler, \underline{N.V. Tran}, C. Vermilion,  J. Walsh) 
}








The internal structure of jets has traditionally been characterized in 
jet shape measurements. A detailed introduction to the current 
theoretical understanding and of the calculations needed for 
observables that probe jet substructure is provided in last year's BOOST 
report~\cite{Altheimer:2012mn}. Here, rather 
than give a comprehensive review of the literature relevant to the myriad of 
developments, we focus on the progress made in the last year in calculations
of jet substructure at hadron colliders. Like the Tevatron experiments 
ATLAS and CMS have performed measurements of the energy flow within the 
jet~\cite{Aad:2011kq,Chatrchyan:2012mec}. Both collaborations
have moreover performed dedicated jet substructure measurements on 
large-R jets that are briefly reviewed before we introduce 
analytical calculations and summarize the status of the two main
approaches. 

\subsection{Jet Substructure Measurements by ATLAS}
\label{subsec:jsATLAS}

The first measurement of jet mass for large-radius jets ($R=1.0$, $1.2$)
and several substructure observables was performed by ATLAS on
data from the 2010 run of the LHC~\cite{ATLAS:2012am}. These early studies
include also a first measurement of the jet mass distribution for
filtered~\cite{Butterworth:2008iy} Cambridge-Aachen jets.
A number of further jet shapes were studied with the same data set
in Reference~\cite{Aad:2012meb}. 
These early studies were crucial to establish
the jet substructure response of the experiment and validate the 
Monte Carlo description of substructure. They are moreover unique, as
the impact of pile-up could be trivially avoided by selecting 
events with a single primary vertex. The results, fully corrected for
detector effects, are available for comparison to calculations.

Since then, the ATLAS experiment has performed a direct and systematic 
comparison of the performance of several grooming algorithms on 
inclusive jet samples, purified samples of 
high-$p_{T}$ $W$ bosons and top quarks, and Monte Carlo simulations of
 boosted $W$ and top-quark signal samples~\cite{Aad:2013gja}. The parameters 
of large-radius ($R=1.0$) trimmed~\cite{Krohn:2009th}, 
pruned~\cite{Ellis:2009me} and mass-drop 
filtered jet algorithms were optimized 
in the context of Standard Model measurements and new physics searches 
using multiple performance measures, including efficiency and jet mass 
resolution. 

For a subset of the jet algorithms tested, dedicated jet energy scale and 
mass scale calibrations were derived and systematic uncertainties evaluated 
for a wide range of jet transverse momenta. Relative systematic uncertainties 
were obtained by comparing ratios of track-based quantities to 
calorimeter-based quantities in the data and MC simulation. \textit{In situ} 
measurements of the mass of jets containing boosted hadronically 
decaying $W$ bosons further constrain the jet mass scale uncertainties for 
this particular class of jets to approximately $\pm1\%$.

\subsection{Jet Substructure Measurements by CMS}
\label{subsec:jsCMS}

The CMS experiment measured jet mass distributions with 
approximately 5~fb$^{-1}$ of data at a center-of-mass 
energy of $\sqrt{s}~=$ 7~TeV~\cite{Chatrchyan:2013rla}.
The measurements were performed in several $p_T$ bins and
for two processes, inclusive jet production and vector boson
production in association with jets. 
For inclusive jet production, the measurement
corresponds to the average jet mass of the highest
two $p_T$ jets. In vector boson plus jet ($V +$ jet) production
the mass of the jet with the highest $p_T$ was measured.
The measurements were performed primarily for jets clustered with the 
anti-$k_t$ algorithm with distance parameter $R=$ 0.7 (AK7).
 The mass of ungroomed, filtered, trimmed, and
pruned jets are presented in bins of pt.
Additional measurements were performed for anti-$k_t$ jets with smaller and
larger radius parameter ($R=$ 0.5, 0.8), after applying 
pruning~\cite{Ellis:2009me} and filtering~\cite{Butterworth:2008iy} 
to the jet, and for Cambridge-Aachen jets
with $R=$0.8 and $R=$ 1.2.  

The jet mass distributions are corrected for detector effects and can
be compared directly with theoretical calculations or simulation models.
The dominant systematic uncertainties are jet energy resolution effects, 
pileup, and parton shower modeling.

The study finds that, for the grooming parameters examined, the pruning 
algorithm is the most aggressive grooming algorithm, leading to the largest
average reduction of the jet mass with respect to the original jet mass.
Due to this fact, CMS also finds that the pruning algorithm reduced 
the pileup dependence of the jet mass the most of the grooming algorithms.

The jet mass distributions are compared against different simulation 
programs: \pythiasix~\cite{Sjostrand:2006za,Sjostrand:2007gs} 
(version 424, tune Z2), 
\hpp~\cite{Corcella:2000bw,Bahr:2008pv} 
(version 2.4.2, tune 23), and \pythiaeight 
(version 145, tune 4C), 
in the case of inclusive jet production.
In general the agreement between simulation and data is reasonable 
although \hpp appears to have the best agreement with the data
for more aggressive grooming algorithms. The $V + $ jet channel appears to 
have better agreement overall than the inclusive jets production 
channel which indicates 
that quark jets are modeled better in simulation.
The largest disagreement with data comes from the low jet mass region, 
which is more affected by pileup and soft QCD effects.

The jet energy scale and jet mass scale of these algorithms were validated
individually. The jet energy scale was investigated in MC simulation, and
was found to agree with the ungroomed energy scale within 3\%, which is
assigned as an additional systematic uncertainty.
The jet mass scale was investigated in a sample of boosted W bosons in a
semileptonic $\ttbar$ sample. The jet mass scale derived from the mass of the
boosted W jet agrees with MC simulation within 1\%, which is also assigned as
a systematic uncertainty.

\subsection{Analytical predictions for jet substructure}

Next-to-leading order (NLO) calculations in the strong coupling constant 
have been performed for multi-jet production, even in association with 
an electro-weak boson. This means that substructure observables, such 
as the jet mass, can be computed to NLO accuracy using 
publicly available codes~\cite{Campbell:2002tg,Nagy:2003tz}.
However, whenever multiple scales, e.g.\ a jet's transverse momentum and 
its mass, are involved in a measurement, the prediction of the observables will 
contain logarithms of ratios of these scales at each order 
in perturbation theory. These logarithms are so important for jet 
shapes that they qualitatively change the shapes as compared to fixed order.
Resummation yields a more efficient organization of the perturbative expansion 
than traditional fixed-order perturbation theory. Accurate calculations
of jet shapes are impossible without resummation. In general one can moreover
interpolate between, or {\it merge}, the resummed and fixed-order result. 

In resummation techniques the perturbative expansion of cross-sections 
for generic observables $v$ is schematically organized in the form\footnote{The 
actual form of \eq{fact} is in general rather complex. For more 
than three hard partons it involves a non-trivial matrix structure
in colour space. 
Moreover, the actual form of the constant terms $g_0(\as)$ depends 
on the flavor of the jet under consideration.}

\begin{align}
& \sigma(v) =\int_0^v dv' \frac{d\sigma}{dv'} = \sum_{\begin{subarray}{c}
\text{partonic}\\ \text{configurations}\\ \delta
      \end{subarray}}
       \sigma_0^{(\delta)} g_0^{(\delta)}(\as) e^{\beta},  
\label{eq:fact}
       \\
      & \beta = {Lg_1^{(\delta)}(\alpha_s L)+g_2^{(\delta)}(\alpha_s L) +\as g_3^{(\delta)}(\alpha_s L) +\dots}
\label{eqresum}
\end{align}
where $\sigma_0=\sum \sigma_0^{(\delta)}$ is the corresponding Born 
cross-section and
$L = \ln v$ is a logarithm of the observable in question\footnote{In the following we concentrate on the case of jet masses with a cut on the jet $p_T$. 
In this case $L=\ln m_J^2/p_T^2$ and $\sigma(v)$ in Eq.~(\ref{eq:fact}) 
is the integrated (cumulative) distribution for $m_J^2< v p_T^2$.}. 

 The notation used in traditional 
fixed-order perturbation theory refers to the lowest-order calculation as 
leading order (LO) and higher-order calculations as next-to-leading order 
(NLO), next-to-next to leading order (NNLO), and so on (with N${^n}$LO 
referring to the $\mathcal{O}(\alpha_s^n)$ correction to the LO result). 
When organized instead in resummed perturbation theory as in \eq{fact}, 
the lowest order, in which only the function $g_1^{(\delta)}$ is retained, 
is referred to as leading-log (LL) approximation. 
Similarly, the inclusion of all $g_i^{(\delta)}$ with $1\leq i\leq k+1$ 
and of $g_0$ up to order $\as^{k-1}$ gives the next$^k$-to-leading log 
approximation to $\ln \sigma$; this corresponds to the resummation of 
all the contributions of the form $\as^n \ln^m N$
with $2(n-k)+1\le m\le 2n$ in the cross section $\sigma$. This can be 
extended to $2(n-k)\le m\le
2n$ by also including the order $\as^k$ contribution to $g_0^{(\delta)}$ .

Typical Monte Carlo event generators such as \pythia, \hpp and
Sherpa~\cite{Gleisberg:2008ta} are correct at LL. 
NLL accuracy has also been achieved for some specific observables, 
but it is difficult to say whether this can be generally obtained. 
Analytic calculations provide a way of obtaining 
precise calculations for jet substructure.
Multiple observables have been resummed (most often at least to NLL 
but not uncommonly to NNLL and as high as  NNNLL accuracy for a few cases) 
and others are actively being studied and calculated in the theory community.

Often for observables of experimental interest, non-global logarithms (NGLs) 
arise~\cite{Dasgupta:2001sh}, in particular whenever a hard boundary in 
phase-space is present (such as a rapidity cut or a geometrical jet boundary). 
These effects enter at NLL level and therefore modify the structure of the 
function $g_2^{(\delta)}$ in \eq{fact}. Until very recently~\cite{Hatta:2013iba}, 
the resummation of NGLs was confined to the limit of large number of 
colours $N_C$~\cite{Dasgupta:2001sh,Dasgupta:2002bw,Banfi:2002hw}. 

Moreover, we should stress that another class of contributions, usually 
referred to as clustering logarithms, affects the $g_2^{(\delta)}$ series 
of \eq{fact} if an algorithm other than anti-$k_t$ is used to define the 
jets~\cite{Banfi:2005gj,Delenda:2006nf}. The analytic structure of these 
clustering effects has been recently explored in 
Ref.~\cite{Kelley:2012zs,Delenda:2012mm} for the case of Cambridge-Aachen 
and $k_t$ algorithms.
 
Furthermore, recent studies have shown that strict collinear factorization is violated if the observable considered is not sufficiently inclusive~\cite{Catani:2011st,Forshaw:2012bi}. As a consequence, coherence-violating (or super-leading) logarithms appear, which further complicate the resummation of certain observables. These contributions affect, for example, non-global dijets observables~\cite{Kyrieleis:2005dt,Forshaw:2008cq} but also some classes of global event shapes~\cite{Banfi:2010xy}.

Of course, to fully compare 
to data one needs to incorporate the effects of hadronization and 
multi-particle interactions (MPI). Progress on this front has also 
been made, both in purely analytical approaches (especially for 
hadronization effects~\cite{Mateu:2012nk}) and in interfacing analytical 
results with parton showers that incorporate these effects.

The two main active approaches to resummation are referred to as traditional 
perturbative QCD resummation (pQCD) and Soft Collinear Effective Theory (SCET).
They describe the same physical effects, which are captured by the 
Eqs.~\ref{eq:fact} and~\ref{eqresum}. However, the techniques employed in pQCD and SCET 
approaches often differ. Calculations in pQCD exploit factorization 
and exponentiation properties of QCD matrix elements and of the 
phase-space associated to the 
observable at hand, in the soft or collinear limits. The SCET approach 
is based on factorization at the operator level and exploits
the renormalization group to resum the logarithms. The two approaches also 
adopt different philosophies for the treatment of NGLs. A more detailed 
description of these differences is given in the next Sections.


\subsection{Resummation in pQCD}
\label{subsec:pQCD}

Jet mass was calculated in pQCD in~\cite{Li:2011hy}. A more 
extensive study can be found in Ref.~\cite{Dasgupta:2012hg} where the jet mass 
distribution for Z+jet and inclusive jet production, with jets defined with the 
anti-$k_t$ algorithm, were calculated at NLL accuracy and matched to LO.
In particular, for the Z+jet case, the jet mass distribution of the 
highest $p_T$ jet was calculated whereas for inclusive jet production, 
essentially the average of jet mass distributions of the two highest 
$p_T$ jets was calculated.
For the Z+jet case, one has to consider soft-wide angle emissions from a
three hard parton ensemble, consisting of the incoming partons and the
outgoing hard parton.
For three or fewer partons, the colour structure is trivial. Dijet production 
on the other hand involves an ensemble of four hard partons and the 
consequent soft wide-angle radiation has a non-trivial colour matrix 
structure. The rank of these matrices grows quickly with the number 
of hard partons, making the calculations for multi jet final states a 
formidable challenge~\footnote{The colour structure of soft gluon 
resummation in a multi jet environment has been studied 
in~\cite{Sjodahl:2008fz,Sjodahl:2009wx} and resummed calculations 
for the case of five hard partons in the context of jet production 
with a central jet veto can be found in 
Refs~\cite{Kyrieleis:2005dt,Forshaw:2008cq,Forshaw:2006fk,DuranDelgado:2011tp}}.


The jet mass is a non-global observable and NGLs of $m_J/p_T$ for jets 
with transverse momentum $p_T$ are induced. Their effect was 
approximated using an analytic formula with coefficients 
fit to a Monte Carlo simulation valid in the large $N_C$ limit, obtained 
by means of a dipole evolution code~\cite{Dasgupta:2001sh}. It was found 
that in inclusive calculations\footnote{We refer to inclusive calculations
if no requirements were made on the number of additional jets in the 
selection of the event.} 
the effects of both the soft wide-angle radiation and the NGLs, both 
of which affect the $g_2^{(\delta)}$ series in \eq{fact}, play a relevant 
role even at relatively small values of jet radius such as $R=0.6$ and 
hence in general cannot be neglected

A restriction on the number of additional jets could be implemented, 
for instance, by vetoing additional jets with $p_T>p_T^\text{cut}$. 
The presence of a jet veto modifies the calculation in several ways. 
First of all, it affects the argument of the non-global logarithms: 
$\ln^n (m_J^2/p_T^2) \to \ln^n (m_J^2/(p_T p_T^{cut}))$. Thus $p_T^\text{cut}$ 
could be in principle used to tame the effect of NGLs. 
However, if the veto scale is chosen such that $p_T^\text{cut}\ll p_T$, 
logarithms of this ratio must be also resummed. Depending on the specific 
details of the definition of the observable, this further resummation can 
be affected by a new class of NGLs~\cite{Banfi:2010pa,KhelifaKerfa:2011zu}.

An obstacle to inclusive predictions in the number of jets is that the constant 
term $g_0^{(\delta)}$ in \eq{fact} receives contributions from higher jet 
topologies that are not related to any Born configurations. For instance, 
the jet mass in the Z+jet process would receive contributions from Z+2jet 
configurations, which are clearly absent in the exclusive case. The full 
determination of the constant term to $\mathcal{O}(\alpha_s)$ and the 
matching to NLO is ongoing.

\subsection{Resummation in SCET}
\label{subsec:SCET}

There have been several recent papers in SCET directly related to 
substructure in hadron collisions\footnote{We consider here only 
research made publicly available at the time of BOOST 2012 or soon after.}. 
Ref.~\cite{Chien:2012ur} discusses the resummation of jet mass by expanding
around the threshold limit, where (nearly) all of the energy goes into the
final state jets. Expanding around the threshold limit has proven effective 
for other observables, see Ref.~\cite{Laenen:1998qw} and references in
Ref.~\cite{Chien:2012ur}.
The large logarithms for jet mass are mainly due
to collinear emission within the jet and soft emission from the recoiling
jet and the beam. These same logarithms are present near threshold and the
threshold limit automatically prevents additional jets from being 
relevant, simplifying the calculation. The study in Ref.~\cite{Chien:2012ur}
performs resummation at the NNLL level, but does not include NGLs. Instead,
their effect is estimated and found to be subdominant in the peak region,
where other effects, such as nonperturvative corrections, are comparable. 
Thus NGLs could be safely ignored where the calculation was most accurate. 

An alternative approach using SCET is found in Ref.~\cite{Jouttenus:2013hs}. 
Beam functions are used to contain the collinear radiation from the beam
remnants. The jet mass distribution 
in Higgs+1jet events is studied via the factorization formula for 1-jettiness, 
that is calculated to NNLL accuracy. Using 1-jettiness, the jet boundaries 
are defined by the distance measure used in 1-jettiness itself, instead of 
a more commonly employed jet algorithm, although generalizations to
arbitrary jet algorithms are possible.

For a single jet in hadron collisions, 1-jettiness can be used as a 
means to separate the in-jet and out-of-jet radiation (see for 
a review the BOOST2011 report\cite{Altheimer:2012mn}). The observable 
studied in Ref.~\cite{Jouttenus:2013hs} is 
separately differential in the jet mass and the beam thrust. 
The in-jet component 
is related to the jet mass, and can be converted directly up to 
corrections that become negligible for higher $p_T$ (up to about 3\% 
for $p_T = 300$~GeV in the peak of the distribution of the in-jet 
contribution to 1-jettiness which is smaller than NNLL uncertainties). 
The beam thrust\footnote{The resummation of beam thrust
is analogous to that of thrust in $e^+e^-$ 
collisions~\cite{Berger:2010xi,Stewart:2010pd,Stewart:2009yx}.} 
is a measure of the out-of-jet contributions, equivalent to a
rapidity-weighted veto scale $p_{\rm cut}$ on extra jets. The calculation
can be made exclusive in the number of jets by making the out-of-jet 
contributions small. 
Where Ref.~\cite{Chien:2012ur} ensures a fixed number
of jets by expanding around the threshold limit, Ref.~\cite{Jouttenus:2013hs}
includes an explicit jet veto scale.

Exclusive calculations in the number of jets avoid some of the 
issues mentioned in \subsec{pQCD}. An important property of 
1-jettiness is that, when considering the sum of the in- and out-of-jet 
contributions, no NGLs are present, and when considering these 
contributions separately, only the ratio $p_{\rm cut} / m_J$ of these two 
scales is non-global. A smart choice of the veto scale may then allow
to minimize the NGL and make the resummation unnecessary.
This corresponds to the NGLs discussed in 
\subsec{pQCD} that are induced in going from the inclusive to the 
exclusive case. These are the only NGLs present; 
the additional NGLs of the measured jet $p_T$ to their mass discussed 
for the observable of \subsec{pQCD} are absent in this case.
By using an exclusive observable, with an explicit veto scale, NGLs
are controlled. For comparison with inclusive jet mass measurements,
such as those discussed in Sections~\ref{subsec:jsATLAS} and~\ref{subsec:jsCMS},
the uncertainty associated with the veto scale can be estimated in
a similar fashion as the NGL estimate in Ref.~\cite{Chien:2012ur}.

It was argued in Ref.~\cite{Jouttenus:2013hs} that the NGLs induced 
by imposing a veto on both the $p_T$ and jet mass are smaller 
than the resummable logarithms of the measured jets over a range of veto 
scales. In contrast,
in the inclusive case the corresponding $p_T$ value that appears in the 
NGLs is of the order of the measured jet $p_T$ (since all values less 
than this are allowed), making it a large scale and the NGLs as large 
as other logarithms. For a fixed veto cut, it was argued that the effect 
of these NGLs (at least of those that enter at the first non-trivial 
order, $\mathcal{O}(\alpha_s^2)$), can be considered small enough to 
justify avoiding resummation for a calculation up to NNLL accuracy 
for $1/\sqrt{8} < m_J^{\rm cut}/p_{\rm cut} < \sqrt{8}$ (cf. 
Ref.~\cite{Hornig:2011iu}) in the peak region where a majority 
of events lie. It is also worth noting that the effect of normalizing 
the distribution by the total rate up to a maximum $m_J^{\rm cut}$ and 
$p_{\rm cut}$ has several advantages and in particular has a smaller 
perturbative uncertainty than the unnormalized distribution, in 
addition to having smaller experimental uncertainties.

We also note that while jet mass is now the most well-understood 
substructure observable, it is also clearly much simpler than the 
more complicated techniques often employed by experimentalists in 
boosted studies. There has also been progress in understanding 
more complicated measurements using SCET, and in particular a 
calculation of the signal distribution in $H \to b \bar{b}$ was 
performed in Ref.~\cite{Feige:2012vc}. While it is probably fair 
to say that our theoretical understanding (or at least the numerical 
accuracy) of such measurements are currently not at the same level 
as that of the jet mass, this is a nice demonstration that reasonably 
accurate calculations of realistic substructure measurements can be 
performed with the current technologies and that it is not 
unreasonable to expect related studies in the near future.

\subsection{Discussion and recommendations for further substructure measurements}
\label{subsec:rec}

We have presented a status report for the two main approaches to 
the resummation of jet substructure observables, with a 
focus on their potential to 
predict the jet invariant mass at hadron colliders. In both approaches
recent work has shown important progress

We hope that providing predictions beyond the accuracy of 
parton showers may help both discovery and 
measurement. Beyond the scope of improving our understanding of QCD, 
gaining intuition for which treatments work best is an important step 
towards adopting such predictions as an alternative to parton showers. 
Non-perturbative corrections like 
hadronization are more complicated at the LHC due to the increased colour 
correlations. Entirely new perturbative and semi-perturbative 
effects such as multiple-particle interactions appear. 
Monte Carlo simulations suggest that these have a significant impact.

The treatments of non-perturbative corrections and NGLs are 
often different in pQCD and SCET~\footnote{We have focused on differences in our
discussion, but typically both communities have the option to adopt the 
treatments commonly employed in the other community. That is, the treatments 
typically utilized are not features inherent to the approach.} and this
leads to slight differences in which measurements are best suited for
comparison to predictions. The first target for the next year should 
be a phenomenological study of the jet mass distribution in Z+jet, for 
which we encourage ATLAS and CMS measurements. Ideally, since the QCD and SCET 
literature have emphasized a difference in preference for inclusive
or exclusive measurements (in the number of jets),
both should be measured to help our understanding of the two 
techniques.

The importance of  boosted-object taggers in searches for new physics 
will increase strongly in the near future in view of the 
higher-energy and higher-luminosity LHC runs. However, the theoretical 
understanding of these tools is in its infancy. Analytic calculations 
must be performed in order to understand the properties of the different 
taggers and establish which theoretical approaches 
(MC, resummation or even fixed order) are needed to accurately compute 
these kind of observables~\footnote{Before the completion of this 
manuscript, two papers appeared~\cite{Dasgupta:2013tia,Dasgupta:2013via} 
which perform analytic resummed calculations for boosted-object methods, 
such as trimming, pruning and mass drop, and energy correlations were 
computed and used for quark and gluon discrimination in 
Ref.~\cite{Larkoski:2013eya}.}.



\section{Monte Carlo Generators for Jet Substructure Observables}
\label{sec:mc}

{\it Section prepared by the Working Group: 'Monte Carlo predictions for jet substructure', \underline{A. Arce}, D. Bjergaard, A. Buckley, M. Campanelli, \underline{D. Kar}, K. Nordstrom.
}

In order to use boosted objects and substructure techniques for measurements and
searches, it is important that Monte Carlo generators describe the jet
substructure with reasonable precision, and that
variations due to the choice of parton shower models and their
parameters are characterized and understood.  
We study jet mass, before and after several jet grooming procedures,
a number of popular jet substructure observables, colour flow and jet charge.
For each of these we compare the predictions of 
several parton shower and hadronisation codes, not only in 
\emph{signal-like} topologies, but also in
background or calibration samples.  


\subsection{Monte Carlo samples and tools}

Three processes in $pp$ collisions are considered
at $\sqrt{s}=7$~TeV: semileptonic \ttbar decays, boosted 
semileptonic \ttbar decays, and $(W^\pm \rightarrow \mu\nu)+$jets.
These processes provide massive jets coming from hadronic decays of a
colour-neutral boson as well as jets from heavy and light quarks. 

Like $Z$+jets, the  ($W\rightarrow \mu\nu)$+jets 
process provides a well-understood source of quarks and gluons, 
and additionally allows an experimentally accessible identification 
(``away-side-tag'') of the charge of the leading jet.  
Assuming that the charge of this jet is opposite to the muon's 
charge leads to the same charge assignment as a conventional parton matching scheme in approximately 70\% of simulated events in leading order Monte Carlo simulation; 
in the remaining 30\% of cases, the recoiling jet matches a (charge-neutral) gluon.

The selection of $t$, $W^\pm$, and quark 
jet candidates for the distributions compared below include event 
topologies that can be realistically collected in the 
LHC experiments, with typical background rejection cuts, so that these studies, 
based on simulation, could be reproduced using LHC data.

\begin{figure*}[t!]
  \centering 
   \includegraphics[width=0.42\textwidth]{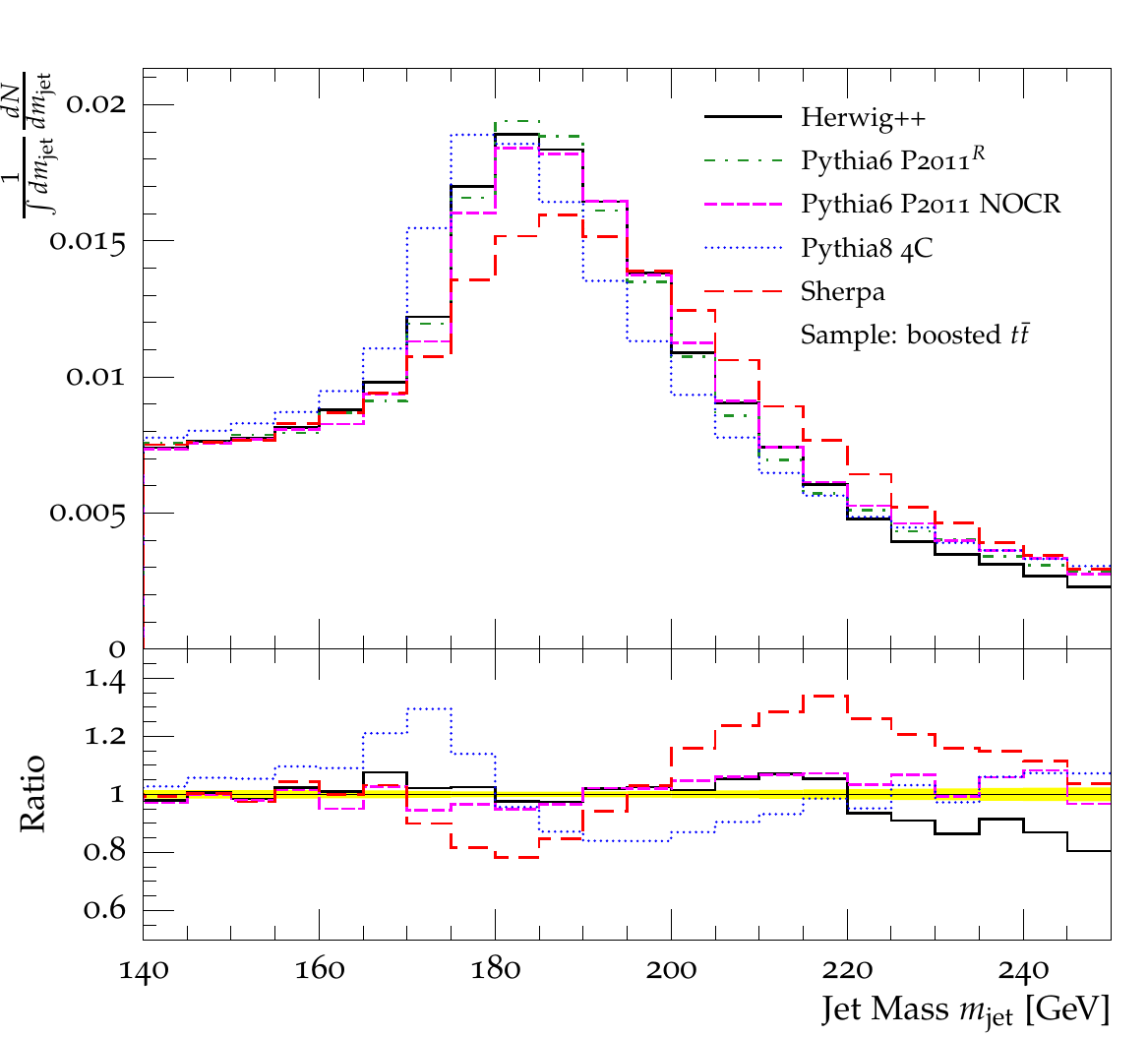}
     \includegraphics[width=0.42\textwidth]{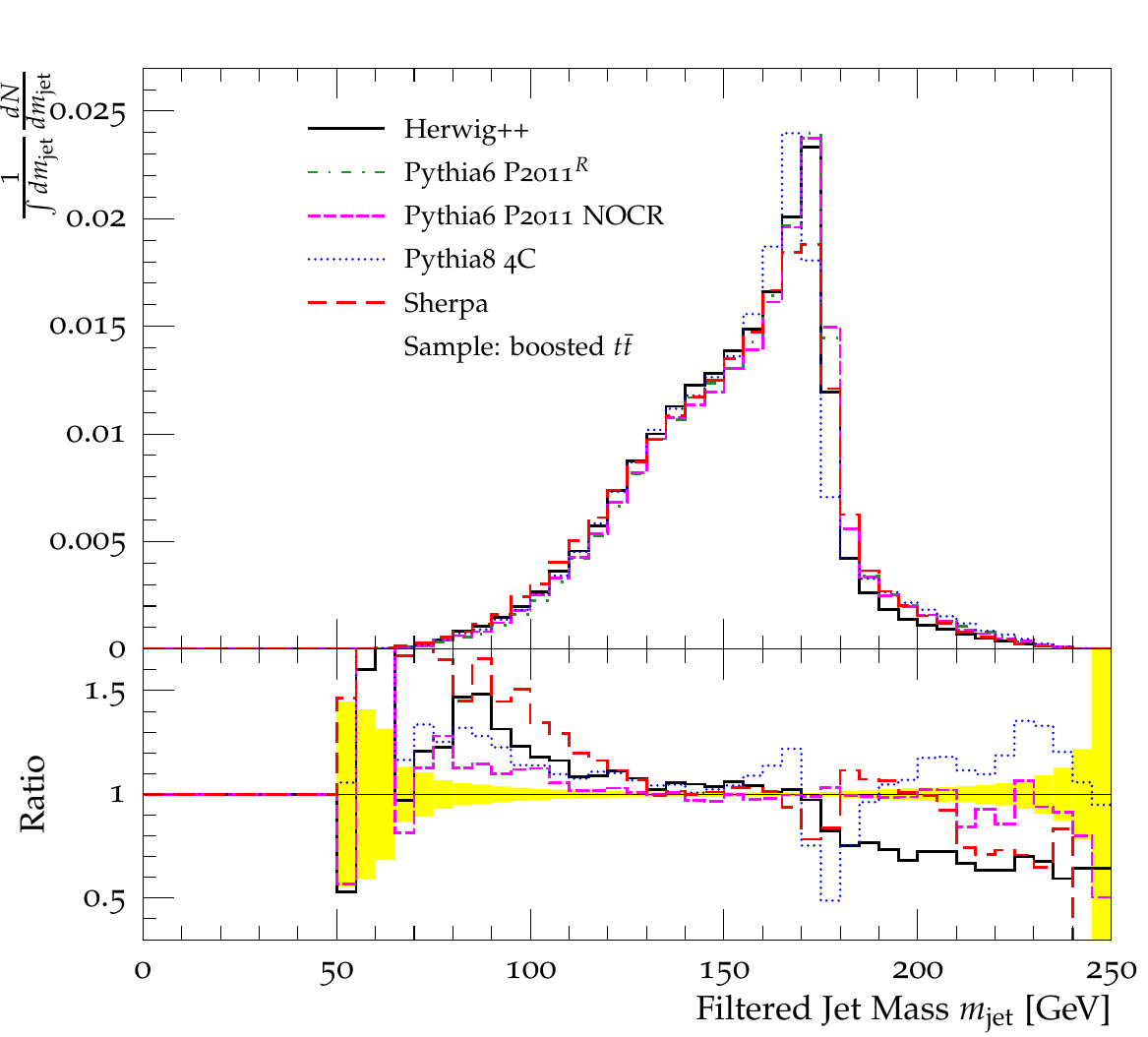} \\
     \includegraphics[width=0.42\textwidth]{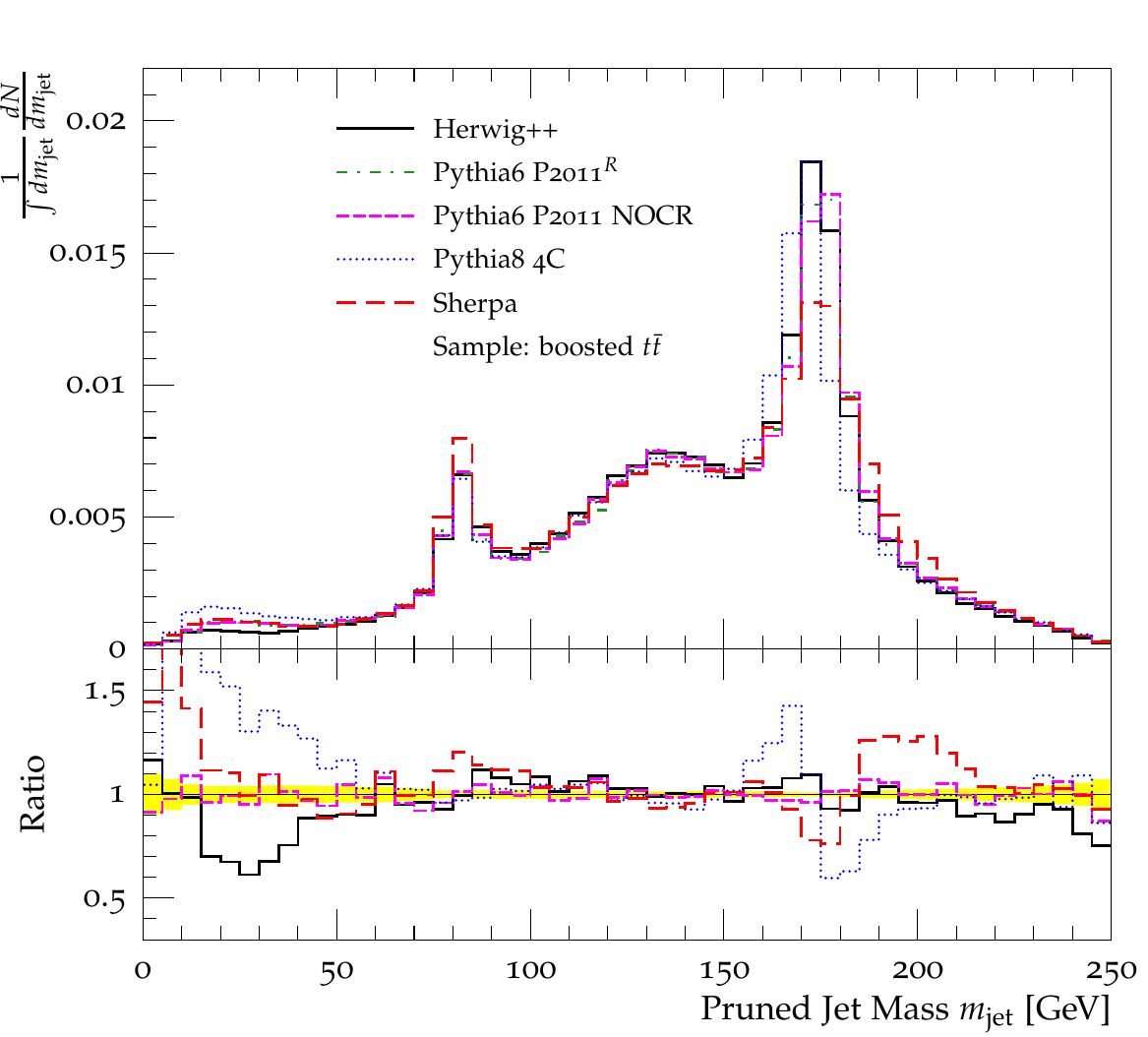}
      \includegraphics[width=0.42\textwidth]{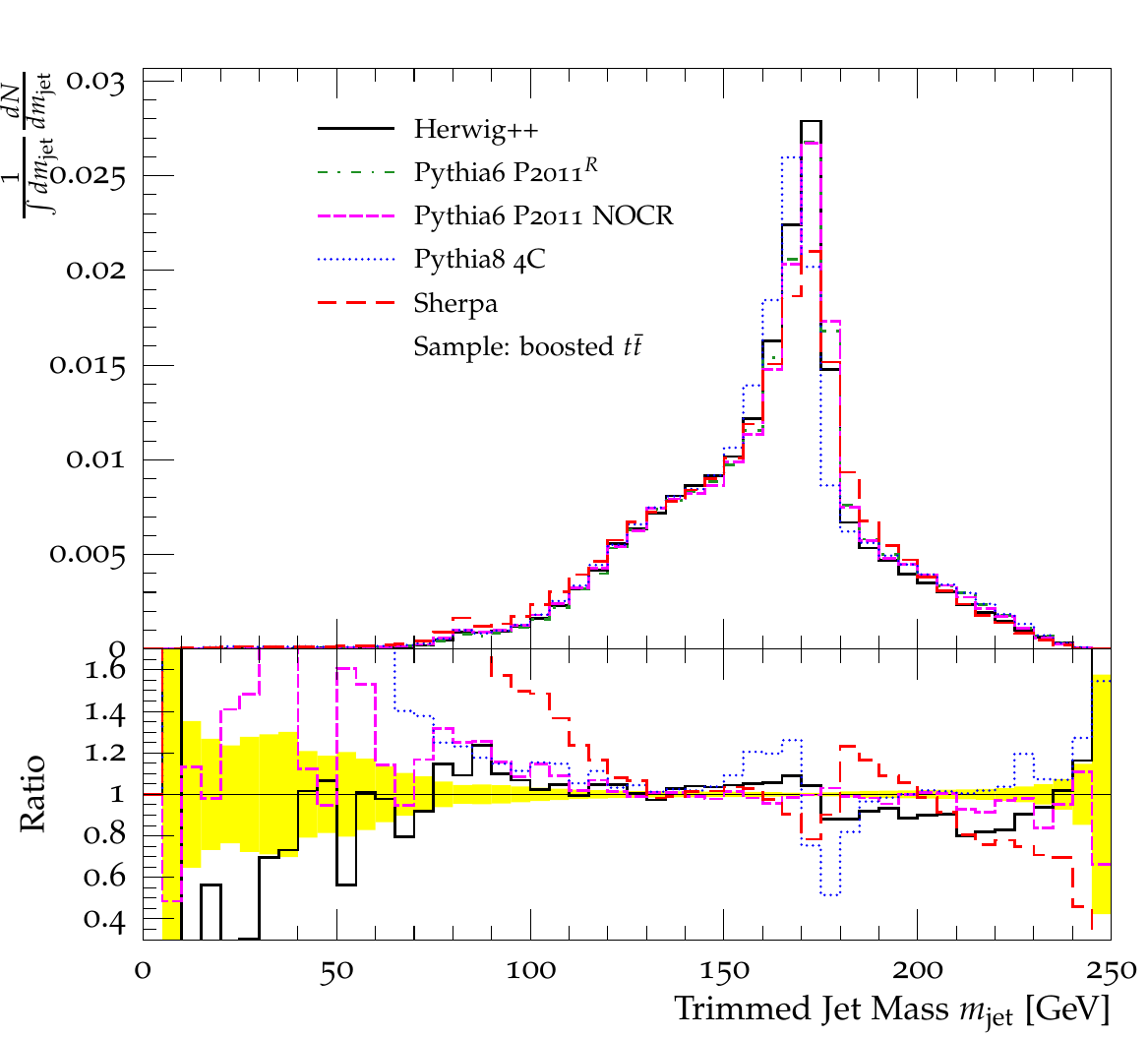}
    \caption[]{
The jet invariant mass distribution for the leading  jet in the boosted 
semileptonic \ttbar{} event sample, before and after jet grooming.}
  \label{fig:jmass}
\end{figure*}

The most commonly used leading order (LO) Monte Carlo simulation codes are
the \textsc{Pythia} and \textsc{Herwig} families. 
Here, predictions from the Perugia 2011~\cite{Skands:2009zm} tune with
CTEQ5L~\cite{Lai:1999wy} parton density function (PDF) and corresponding 
NOCR tunes of \textsc{pythia6}~\cite{Sjostrand:2000wi,Sjostrand:2006za}
(version 6.426), tune 4C~\cite{Corke:2010yf} with CTEQ6L1 PDF~\cite{cteq6} 
of the newer \texttt{C++} \textsc{Pythia8} generator~\cite{Sjostrand:2007gs}
(version 8.170), and the LHC-UE-EE-4~\cite{Gieseke:2012ft} tune of
\hpp~\cite{Bahr:2008pv,Gieseke:2011na} (version 2.6.1) with CTEQ6L1 
PDF are compared. The default parameter tune of the
next-to-leading order (NLO) parton shower model implemented in
\textsc{Sherpa}~\cite{Gleisberg:2008ta} (version 1.4.2) with CT10 
PDF~\cite{Lai:2010vv} is also included in comparisons. 
The \textsc{Pythia6} generator with the Perugia2011 tune is taken as a 
reference in all comparisons. 
For each generator, tune and process 1 million proton-proton events at $\sqrt{s}=$ 7~\tev{} are produced.

The analysis relies on the FastJet 3.0.3 package~\cite{Cacciari:2011ma,Cacciari:2008gn} 
and Rivet analysis framework~\cite{Buckley:2010ar}. All 
analysis routines are available on the conference web page~\cite{rivetroutines}.
In the boosted semileptonic \ttbar analysis, 
large-radius jets were formed using the anti-$k_t$ 
algorithm~\cite{Cacciari:2008gp} with a radius parameter of $1.2$
using all stable particles within pseudorapidity $|\eta| < 4$.
The jets are selected if they passed 
the following cuts: $p_T^{\text{jet}} > 350$ GeV, 140 GeV
$ < m^{\text{jet}} < 250$ GeV. 
Only the leading and subleading jets were selected if more than two 
jets passed the cuts.
The subjets were formed using the Cambridge-Aachen 
algorithm~\cite{Dokshitzer:1997in,Wobisch:1998wt} with radius $0.3$.

\subsection{Jet mass}

The jet mass distribution for the leading jet in the boosted semi-leptonic
\ttbar{} sample is shown in Fig.~\ref{fig:jmass}. The  
parton shower models in \textsc{Pythia6}, \textsc{Pythia8}, \textsc{Herwig++} 
and \textsc{Sherpa} yield significantly different predictions. 
Important differences
are observed in the location and shape of the top quark mass peak.
The largest deviations of the normalized cross section 
in a given jet mass bin amount to approximately 20\%. 
Much better agreement is obtained for predictions with different tunes of a 
single generator.

The effect of different grooming techniques on jet mass is also shown in 
Fig.~\ref{fig:jmass}. For filtering, three hardest subjets with $R^{sub} =0.3$ are used.
The trimming uses all subjets over $3\%$ of $p_T^{jet}$ and $R^{sub} = 0.3$.
For pruning, $z = 0.1$ and $D = m^{jet} / p_T^{jet}$ is used.
As expected, a much narrower top quark mass peak 
is obtained, with a particularly strong reduction of the high-mass tail. 
The grooming procedure improves the agreement among the 
different Monte Carlo tools, as expected from previous Monte Carlo studies
with a more limited set of generators~\cite{Abdesselam:2010pt} and
comparison with data~\cite{ATLAS:2012am}.

\begin{figure*}[t!]
  \centering
\includegraphics[width=0.42\textwidth]{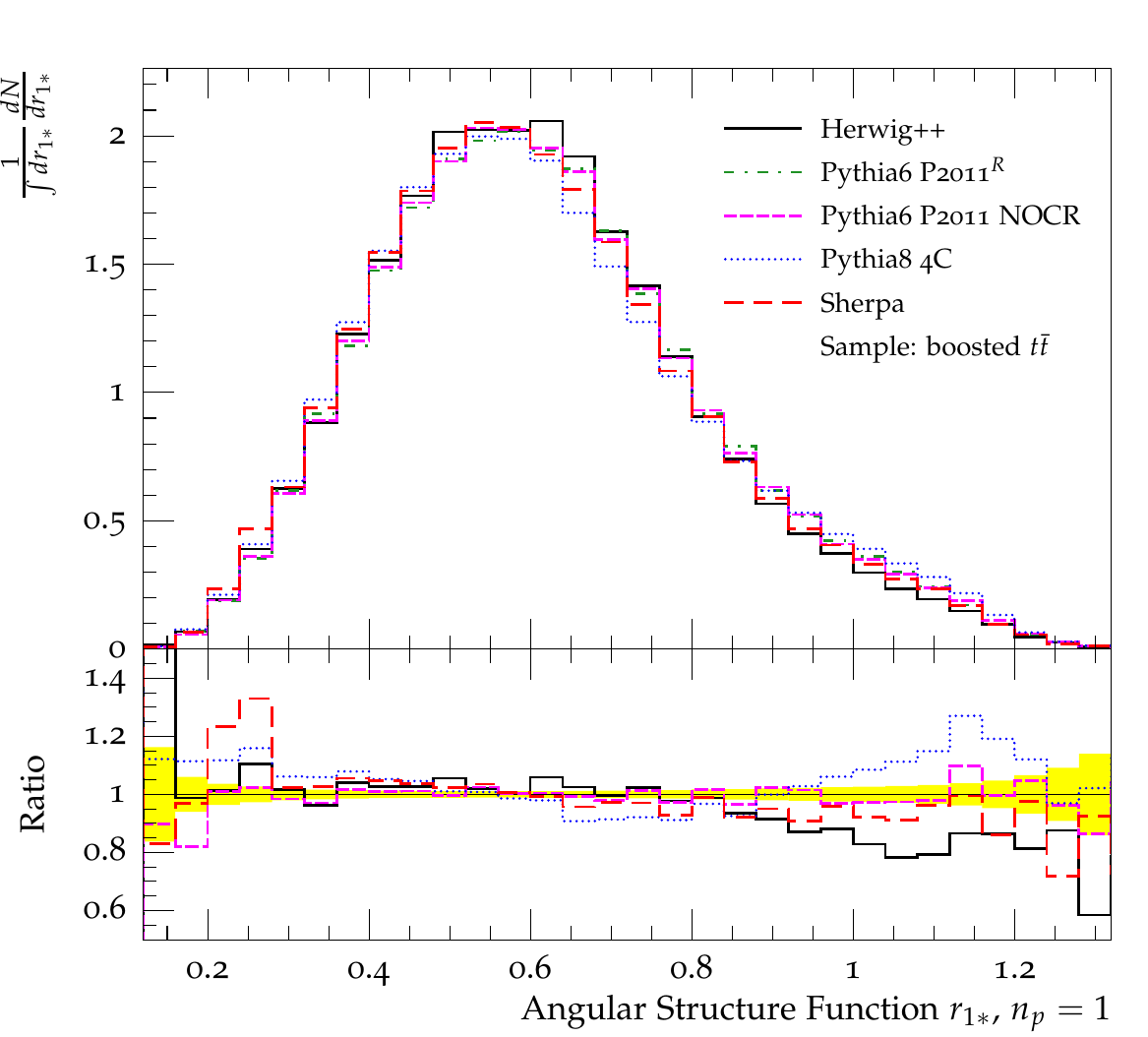}
\includegraphics[width=0.42\textwidth]{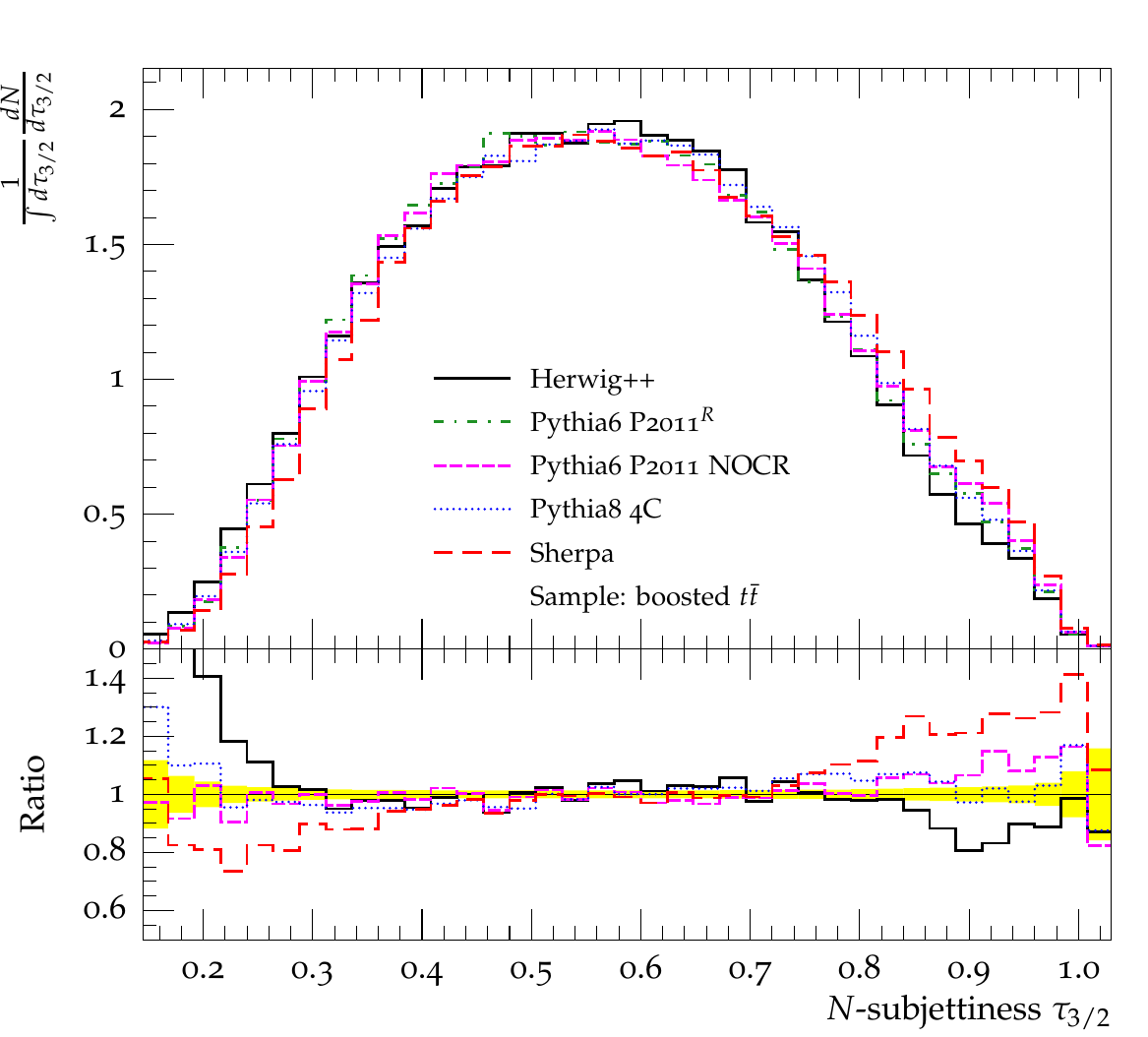} \\
\includegraphics[width=0.42\textwidth]{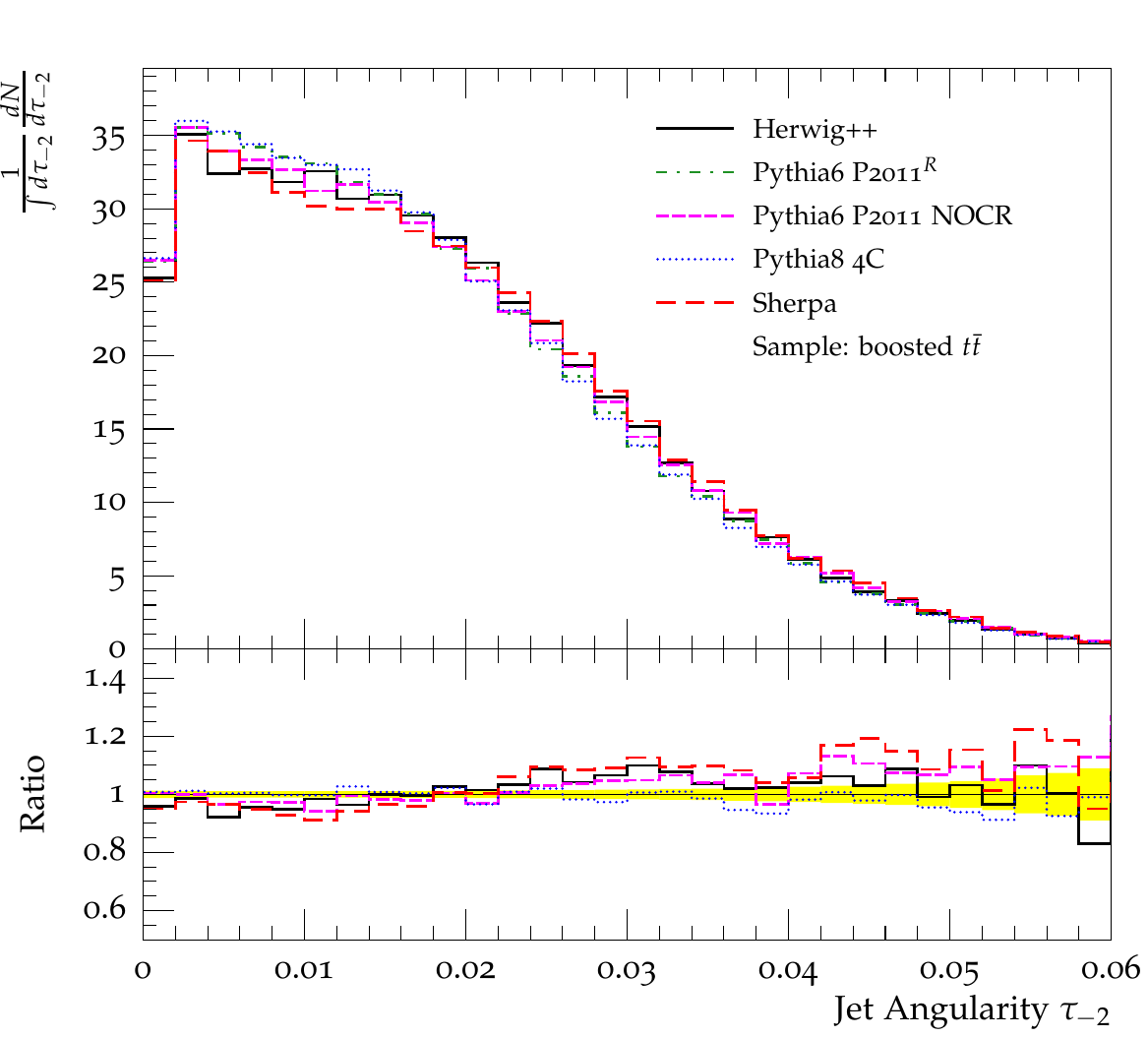}
\includegraphics[width=0.42\textwidth]{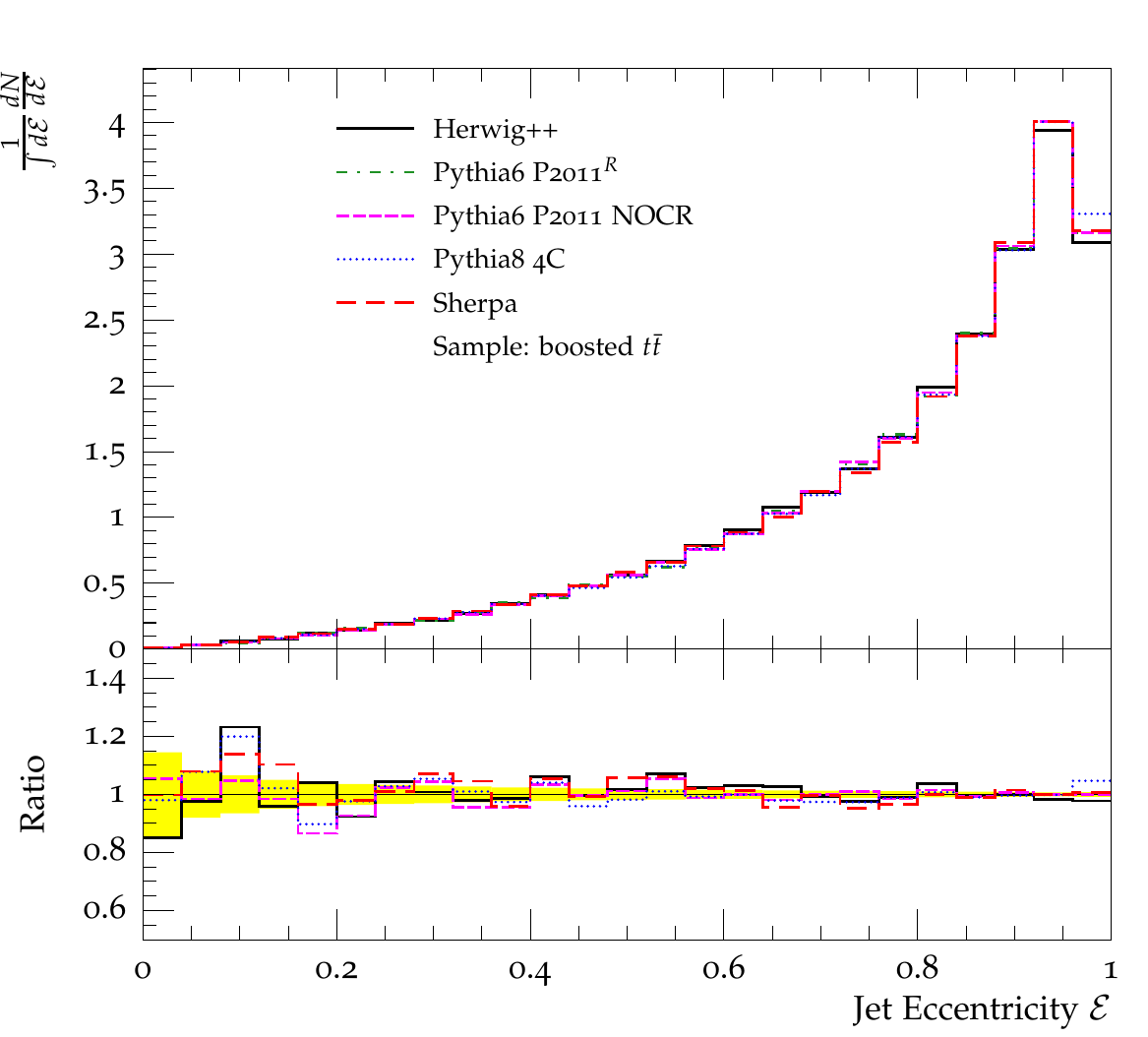} \\
         \caption[]{The distribution 
         of four different measures of jet substructure
         for leading jet of a boosted semileptonic top sample. 
         The C-A algorithm is used in reclustering, as mentioned in the text.}
  \label{fig:substructureobservables}
\end{figure*}

\begin{figure*}[t!]
  \centering
     \includegraphics[width=0.42\textwidth]{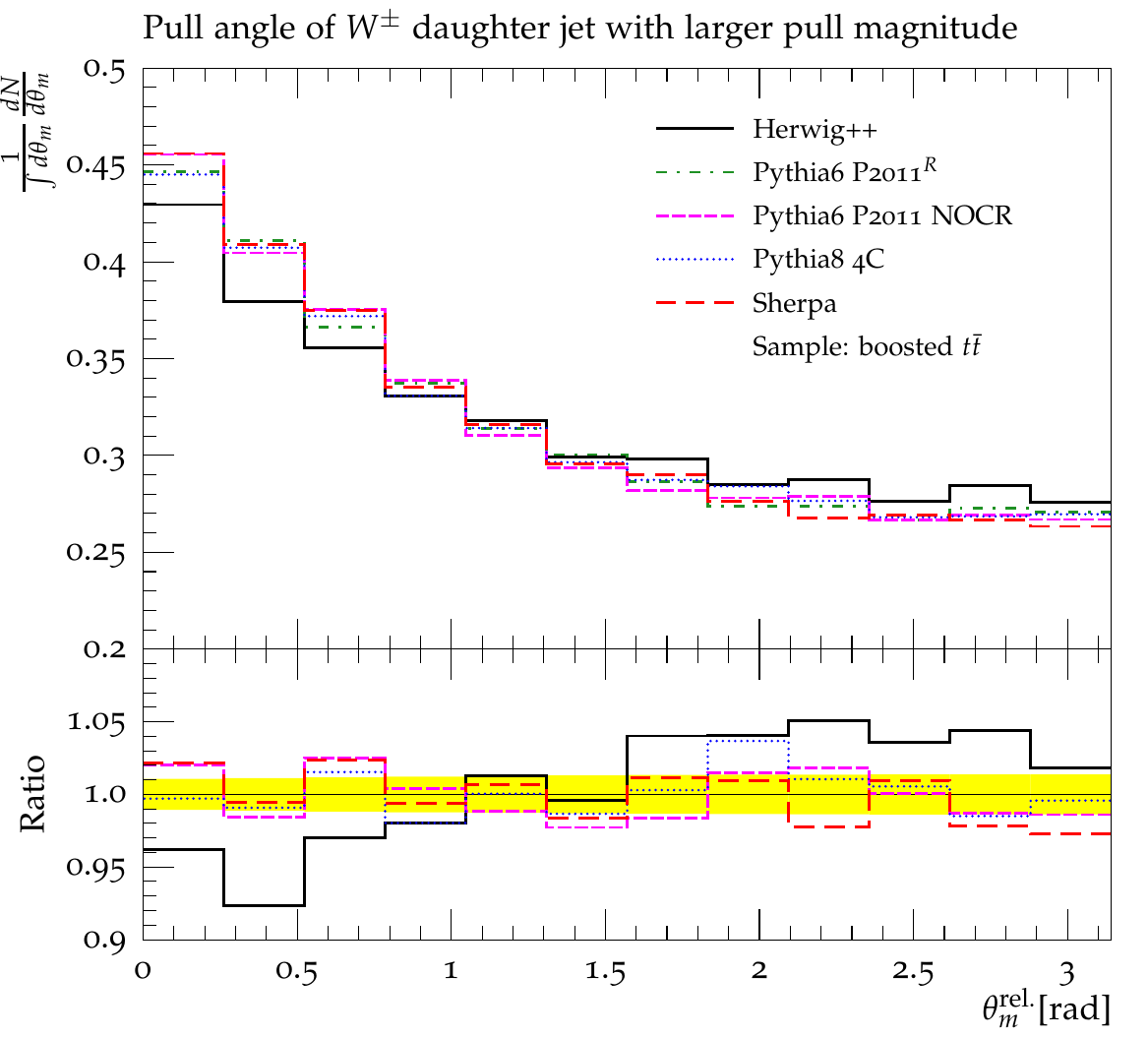} 
      \includegraphics[width=0.42\textwidth]{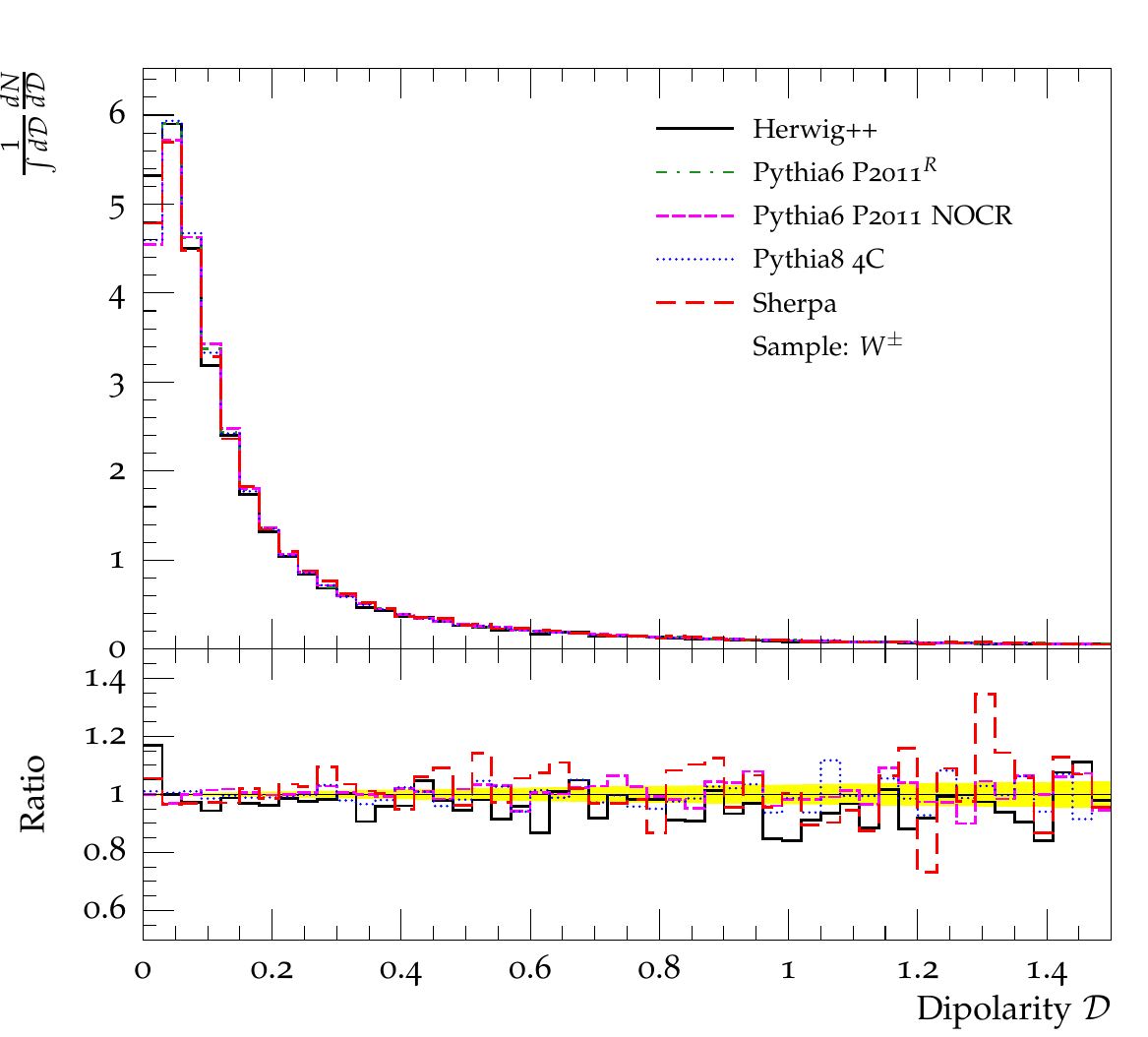}
      \includegraphics[width=0.42\textwidth]{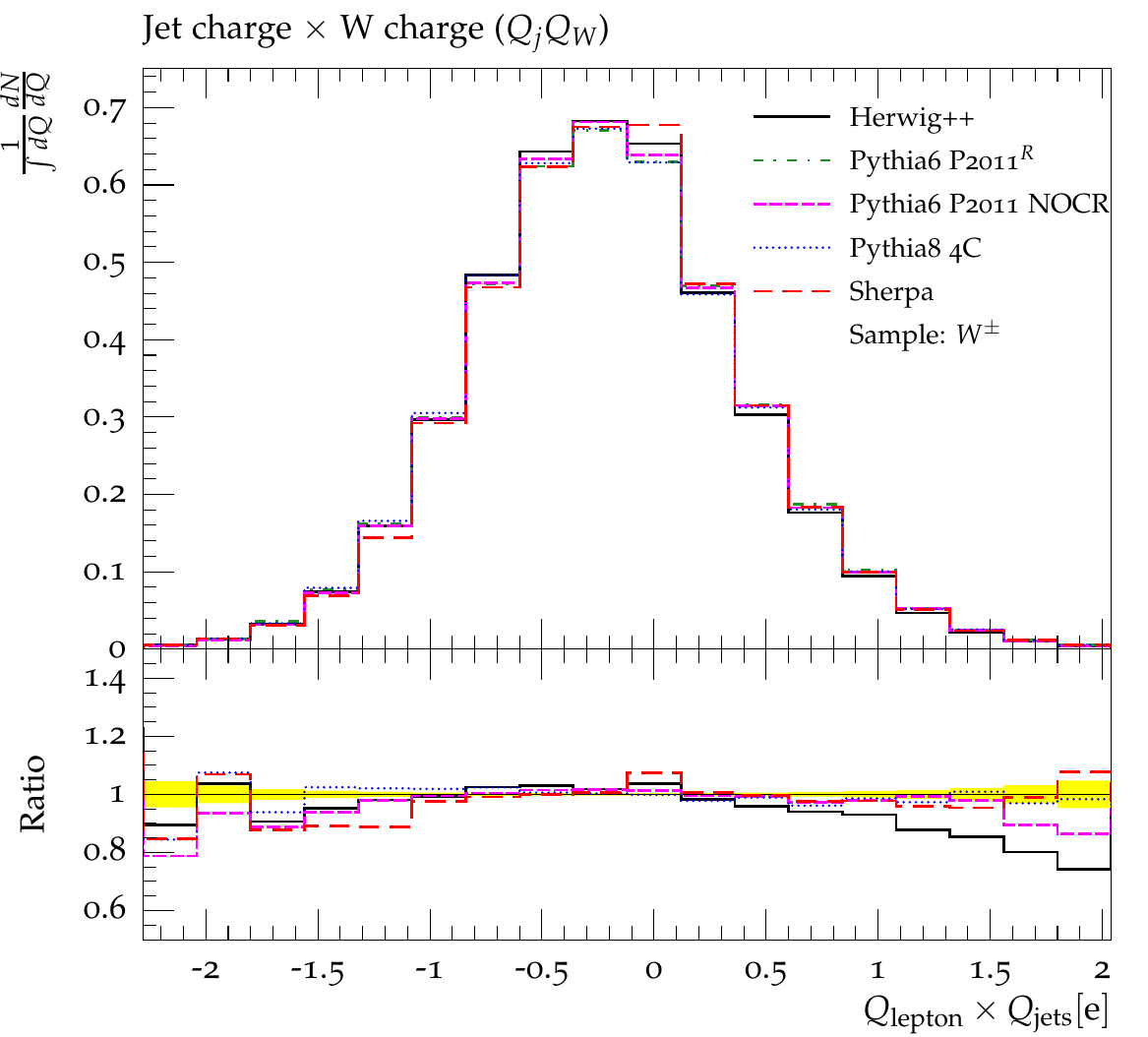}
      \includegraphics[width=0.42\textwidth]{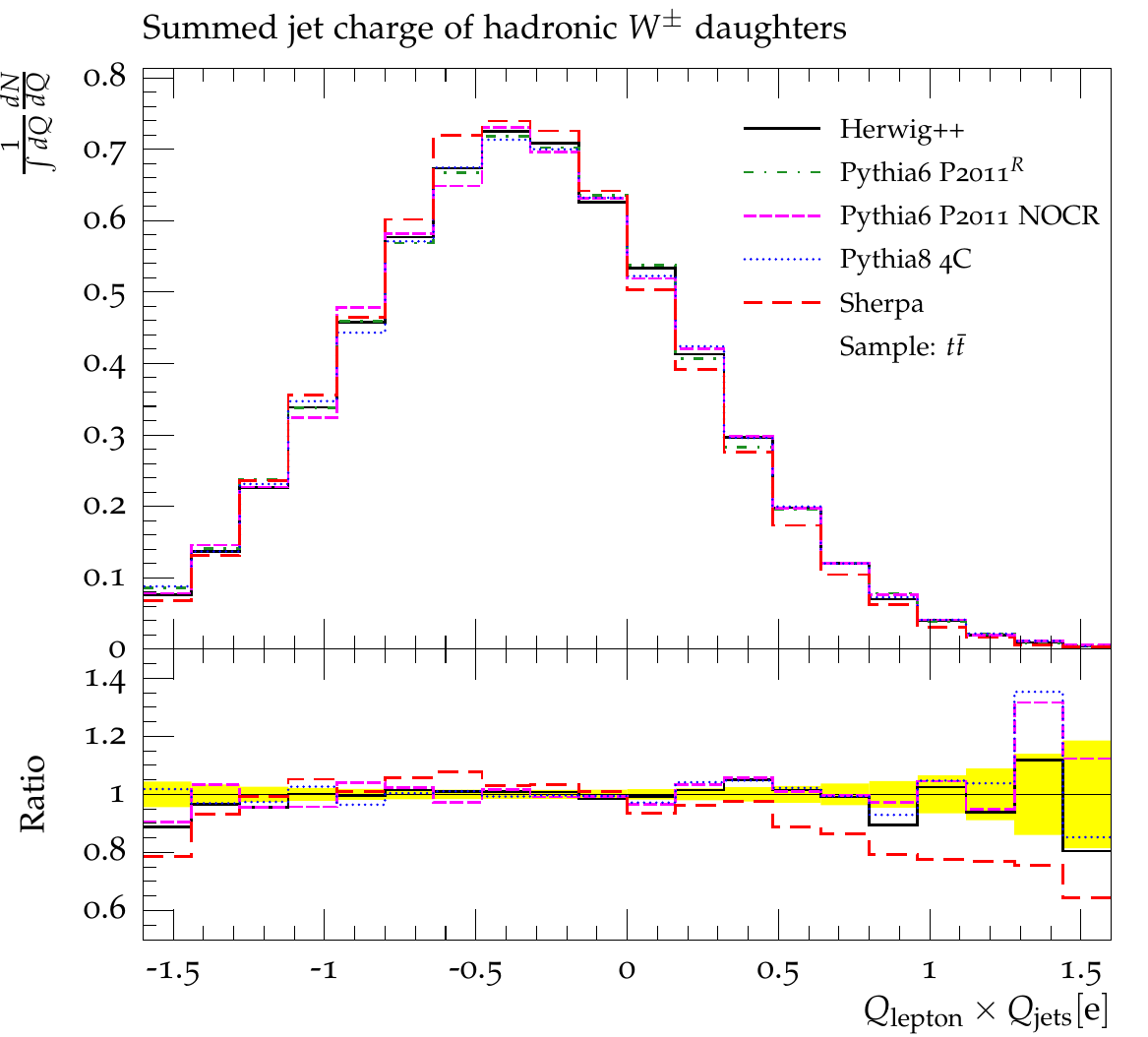}
 \caption[]{
 Upper row: Comparison of colour-flow observables: pull angle of leading jet 
 attributed to the hadronic W decay in \ttbar{} events, and 
 dipolarity of leading jet produced in association with a leptonically decaying W.
 Lower row: Comparison of jet charge observables  ($\kappa=0.3$): 
 charge observable for leading jet produced in association with a 
 leptonically decaying W (left panel), and sum of jet charge observables for the 
 two jets attributed to the hadronic W  decay in \ttbar{} events (right panel).
}
  \label{fig:cflow}
\end{figure*}

\subsection{Jet substructure observables}
We investigate the spread among generators for a number of other 
substructure observables on the market:

\begin{itemize}

\item The Angular Correlation Function~\cite{Jankowiak:2011qa}
measures the $\Delta R$ scale of a jet's radiation. It is defined as:
\[ \mathcal{G}(R) = \frac{1}{\sum p_{T,i} p_{T,j} \Delta R_{i,j}^2} \sum p_{T,i} p_{T,j} \Delta R_{i,j}^2 \Theta(R - \Delta R_{i,j}) \]
where the sum runs over all pairs of particles in the jet, 
and $\Theta(x)$ is the Heaviside step function. 
The Angular Structure Function is defined as the following derivative:
\[ \Delta \mathcal{G}(R) = \frac{ d \hspace{0.1cm} \text{log} \hspace{0.1cm} \mathcal{G}(R) }{d \hspace{0.1cm}\text{log} \hspace{0.1cm} R} \]
Peaks in $\Delta \mathcal{G}(R)$~\footnote{
In this analysis the derivatives are smoothed using a 
Gaussian in the numerator and an error function in the denominator, both 
with $\sigma = 0.06$}
can then be found which 
correspond to $\Delta R$ scales with excess radiation in the jet. 
The variable $r_{1*}$ is the point in the $dR$-spectrum that the first peak in the angular structure 
function appears at,  and $n_p$ is the total number of peaks in the jet's angular structure function.
The prominence $h$ of the highest peak is 
defined as its height. The prominence of any lower peak 
is defined as the minimum vertical descent that is required in descending 
from that peak before ascending a higher, neighboring peak.
A prominence of $h > 4$ for peaks in the angular 
structure function is required and the partial mass and $\Delta R$ scale
of the most prominent peaks are retained.

\item $N$-subjettiness~\cite{Thaler:2010tr,Thaler:2011gf} measures how much 
of a jet's radiation is aligned along 
$N$ subjet axes in the $y - \phi$ plane. It is defined as:
\[ \tau_N = \frac{1}{\sum\limits_k p_{T,k} R_{\text{jet}}^\beta } \sum\limits_k p_{T,k} \hspace{0.1cm} \text{min}(\Delta R_{1,k}^\beta, \Delta R_{2_k}^\beta, ...) \]
where $\Delta R_{n,k}$ is the distance from $k$ to the $n$th subjet axis in the $y - \phi$ plane, 
$R_{\text{jet}}$ is the radius used for clustering the original jet, 
and $\beta$ is an angular weighting exponent\footnote{To improve the performance of $N$-subjettiness it is possible to use a k-means clustering 
algorithm to find (locally) optimal locations for the subjet axes. 
In this analysis  $\beta = 1$ is used to find the subjet axes by reclustering with the $k_t$ algorithm.
The $k$-means clustering algorithm is run once,  
as with this angular weighting exponent it finds a local minimum immediately. 
No attempt is made to find the global minimum.}.

\item Angularity~\cite{Berger:2003iw} introduces an adjustable parameter
$a$ that interpolates between the well-known event shapes thrust 
and jet broadening. Jet angularity is an IRC safe variable (for $a<$ 2) 
that can be used to separate multijet background from jets
containing boosted objects~\cite{Almeida:2008yp}.
It is defined as:
\[ \tau_a = \frac{1}{m_{jet}} \sum\limits_{i\in jets} \omega_i \sin ^a \theta_i (1 - \cos \theta_i)^{1-a}  \]
where $\omega_i$ is the energy of a constituent of the jet.

\item 
Eccentricity~\cite{Chekanov:2010vc} of jets is defined by 
$1-v_{\text{max}}/v_{\text{min}}$, 
where $v_{max}$ and $v_{min}$ are the maximum and minimum values
of the  variances of jet constituents along the principle and minor 
axes\footnote{Eccentricity is strongly correlated with the 
planar flow, and it is a measure of jet elongation ranging
from 0 for perfectly circularly symmetric jet shapes to 1 for 
infinitely elongated jet shapes. This is primarily useful for
identifying high $p_T$ merged jets.}.

\end{itemize}

Most models predict very similar behavior for angularity, eccentricity
and the $\Delta R$ scale of the peak in the $n_p=1$ bin for the angular structure function.
Deviations are typically below 10\% for these observables.
The harder jet mass distribution in \textsc{SHERPA} and the softer spectrum
in \textsc{Pythia8} are reflected in the edges of the $\tau_{3/2}$ 
distribution.

\subsection{Colour flow}

Colour flow observables offer a complimentary way to probe boosted event 
topologies. Pull~\cite{Gallicchio:2010sw} is a $p_T$-weighted vector in 
$\eta-\phi$ space that is constructed so as to point from a given jet to 
its colour-connected partner(s). The pull is measured with 
respect to the other $W^\pm$ daughter jet. 
The $W$-boson is selected kinematically in $4$-jet events with $2$ 
$b$-quarks, and flavors are labelled using the highest $p_T$ cone.
In Fig.~\ref{fig:cflow}, the top left plot 
shows this variable for a background-like distribution. 
The comparisons demonstrate 
that Herwig produces a different colour flow structure.

Dipolarity~\cite{Hook:2011cq} can distinguish whether a pair of subjets arises
from a colour singlet source. In the top right plot of Fig.~\ref{fig:cflow}, 
the dipolarity predictions are seen to be similar for all models considered.

\subsection{Jet charge}

Jet charge~\cite{Feynman1978,Waalewijn2012,Krohn2012} is constructed 
in an attempt to associate a jet-based observable to the charge of the 
originating hard parton. The $p_T$-weighted jet charge
\[
Q_j = \frac{1}{{p_T}_j^\kappa}\sum_{i\in T} q_i\times (p_T^i)^\kappa \]
is shown with $\kappa=0.3$ in Fig.~\ref{fig:cflow}, using anti-kt $0.6$ jets.
The comparison displays the most relevant distributions for typical quark tagging and boson tagging analyses.
Different MC models are seen to have very similar predictions for this observable too.

\subsection{Summary}

We have prepared the Rivet routines to evaluate the predictions of 
Monte Carlo generators for the internal structure of large 
area jets. The normalized predictions from several mainstream
Monte Carlo models are compared. Several aspects of jet substructure 
are evaluated, from basic jet invariant mass to colour flow observables 
and jet charge.

We find that for jet mass large variations are observed 
between the various MC models. 
However, for groomed jets the deviations between different model predictions 
are smaller. The differences between several recent tunes of the 
\textsc{Pythia} generator are much smaller.
The MC model predictions are similar for $N$-subjettiness, angularity 
and eccentricity. The \hpp model gives different predictions than other models 
for colour flow observables, but since the implementation of colour 
connection in \hpp model is very recent, this may lead to improvement 
of the model.

\section{The impact of multiple proton-proton collisions on jet reconstruction}
\label{sec:pileup1}


{\it Section prepared by the Working Group: 'Jet substructure performance at high luminosity', P. Loch, D. Miller, K. Mishra,  P. Nef, \underline{A. Schwartzman}, \underline{G. Soyez}.
}






The first LHC analyses exploring the experimental response to jet substructure
demonstrated that the highly granular ATLAS and CMS detectors can yield 
excellent performance. They also confirmed the susceptibility of 
the invariant mass of large-area jets to the energy flow from the 
additional proton-proton interactions that occur each bunch crossing. 
And, finally, they provided a first hint that jet grooming could be a 
powerful tool to mitigate the impact of pile-up. Since then, the LHC
collaborations have gained extensive experience in techniques 
to correct for the impact of pile-up on jets. In this Section 
these tools are deployed in an extreme pile-up environment. 
We simulate \pu{} levels as high as $\axing = 200$, such as may be
expected in a future high-luminosity phase of the LHC. We evaluate
the impact on jet reconstruction, with a focus on the (substructure) 
performance.

\subsection{Pile-up}
Each LHC bunch crossing gives rise to a number of \pp{} collisions and typically the hard scattering (signal) interaction is accompanied by several additional \pu{} \pp{} collisions. The total \pp{} cross-section is about $\sigma_{\mathrm{tot}} = 98$ mb (inelastic $\sigma_{\mathrm{inel}} = 72.9$ mb) at \sqrts{7} \cite{Antchev:2013iaa}, and even slightly higher at \sqrts{8}{} in 2012. With a peak instantaneous luminosity of about $7.7 \times 10^{33}$~cm$^{-2}$~s$^{-1}$ in 2012, the resulting average number of \pu{} collisions reached $\axing = 20$ at the highest intensities. The 2012 data set has a rather flat $\mu$ distribution extending from $\mu = 5$ to $\mu = 35$. In future \LHC{} running even higher \axing{} are expected. 

\PU{} manifests itself mostly in additional hadronic transverse momentum 
flow, which is generated by overlaid and statistically independent, 
predominantly soft \pp{} collisions that we refer to as 
``minimum bias'' (MB). This diffuse 
transverse energy emission interferes with the signal of hard scattering 
final state objects like particles and particle jets, and typically 
requires corrections, in particular for particle jets.  In addition,
 it can generate particle jets (\emph{\pujet s}) either from any of 
the individual \MB{} collisions (\emph{\qcdjet s}), or by stochastically 
forming jets in the high density particle flow generated by the 
multitude of them (\emph{\stojet s}).  

\subsection{Monte Carlo event generation} \label{subsec:mc} 

We model the \pu{} with \MB{} collisions at \sqrts{8} and
a bunch spacing of 50 ns, generated with the \pythia{} Monte Carlo
(MC) generator~\cite{pythia,pythia8}, with its 
{\sc 4C} tune  \cite{Corke:2010yf}. 
All inelastic, single diffractive, and double diffractive processes are 
included, with the default fractions as provided by \pythia (tune 4C). 

Overall $100\times 10^{6}$ \MB{} events are available for 
pile-up simulation. The corresponding data are generated in samples of 
$25000$ \MB{} collisions, with the largest possibly statistical 
independence between samples, including new random seeds for each sample. 
To model pile-up for each signal interaction, the stable 
particles\footnote{A particle is considered stable if its lifetime $\tau$ 
in the laboratory frame of reference passes $c\tau > 10$~mm.} generated 
in a number $\mu$ of \MB{} collisions, with $\mu$ being sampled from a 
Poisson distribution around the chosen \axing, are added to the final 
state stable particles from the signal. This is done dynamically by an 
event builder in the analysis software, and is thus not part of the 
signal or \MB{} event production. All analysis is then performed on 
the merged list of stable particles to model one full collision event 
at the \LHC. 


The example signal chosen for the Monte Carlo simulation based studies
presented in this Section is the decay of a possible heavy \Zprime{}
boson with a chosen $\ZprimeM = 1.5$~\tev{} to a (boosted) top
quark pair, at \sqrts{8}. The top- and anti-top-quarks then decay 
fully hadronically ($t \to \Wboson b \to jj\,\bjet$) or 
semi-leptonically ($t \to \Wboson b \to \ell \nu\,\bjet$). 
The \pythia{} generator \cite{pythia,pythia8} is used to generate the 
signal samples. The soft physics modeling parameters in both cases are 
from the pre-LHC-data tune 4C  \cite{Corke:2010yf}.  
The pile-up is simulated by overlaying generated minimum bias
\pp{} interactions at \sqrts{8} using Poisson distributions with
averages $\axing = \{ 30, 60, 100, 200 \}$, respectively, thus
focusing on the exploration of future high intensity scenarios at
\LHC.

All analysis utilizes the tools available in the \FJ{} \cite{Cacciari:2011ma} 
package for jet finding and jet sub-structure analysis. The larger jets used 
to analyze the final state are reconstructed with the \antikt{} 
algorithm~\cite{Cacciari:2008gp} with  $R = 1.0$, to assure that most of 
the final state top-quark decays can be collected into one jet. This 
corresponds to top-quarks generated with $\pT \gtrsim 400$~\GeV. 
The configurations for jet grooming are discussed 
in Section \ref{sec:grooming}.

\subsection{Investigating jets from \pu} \label{sec:investigation}

Stable particles emerging from the simulated \pp{} collisions are clustered into \antikt{} jets \cite{Cacciari:2008gp} with
a radial distance parameter $R=0.4$, using the \FJ{} \cite{Cacciari:2011ma} implementation:
\begin{description}
    \item{\bf Truth jets} are obtained by clustering all stable particles from a given individual \MB{} interactions. For an event containing
        $\mu$ pileup interactions, jet finding is therefore executed $\mu$ times.
	The resulting truth jets are required to have $\pT \geq 5$~\GeV. 
    \item{\bf Pileup jets} are obtained by clustering the stable particles from all \MB{} 
            interactions forming the \pu{} event. They are subjected to the kinematic cuts described below. 
\end{description}
Jets with rapidity $|y|<2$ are accepted.

The contribution of \pu{} to jets can be corrected using the jet area based method in Ref.~\cite{Cacciari:2007fd}. It employs the median transverse momentum density $\rho$, which here is determined using \kT{} jets with $R = 0.4$ within $|y| < 2$.   To evaluate the effect of this correction, the transverse momentum ratio \Rpt{} is introduced as
\begin{equation}
\Rpt =\dfrac{\ptmatch}{\pT - \rho A} = \dfrac{\ptmatch}{\ptcorr}.
\label{eq:match}
\end{equation} 
Here $A$ is the catchment area \cite{Cacciari:2008gn} of the \pujet, and \ptmatch{} is the matching truth jet \pT. The matching criterion is similar to the one suggested in Ref.~\cite{Cacciari:2010te}, where the truth jet matching uses the constituents shared between the truth jet and the \pujet. The jets are considered matched if  the fraction of constituents of the truth jet that are also contained in the \pujet{} contribute to at least 50\% of the truth jet \pT. 
In the following, \pujet s are only considered if their corrected transverse momentum is $\ptcorr\geq 20$~\GeV, and they are matched to at least one truth jet. 


\begin{figure}[htbp!]
  \begin{center} 
    \includegraphics[width=0.46\textwidth]{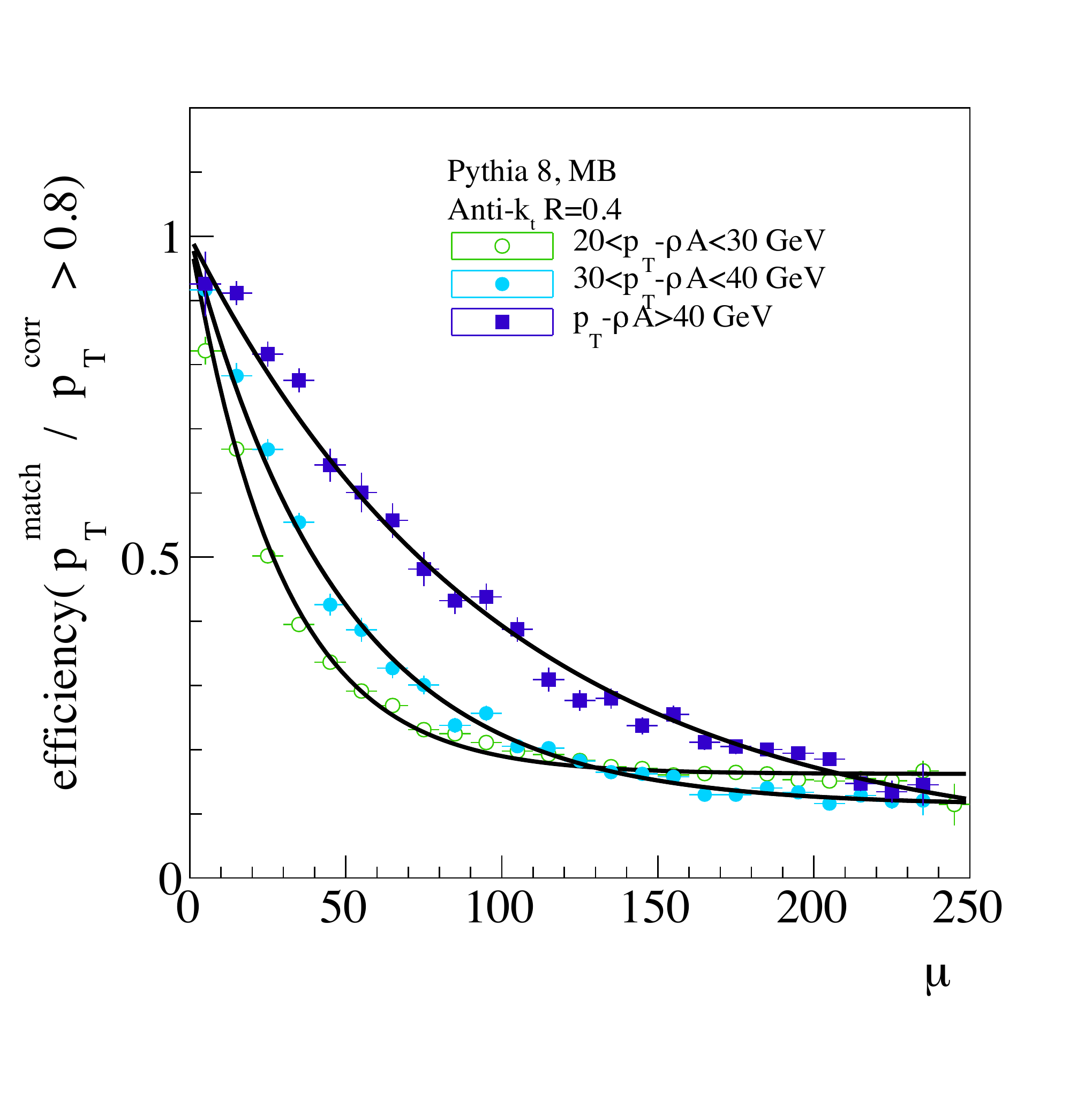}
    \caption[]{The fraction of \pujet s with $\Rpt > 0.8$ (QCD-like) as a function of the number of minimum-bias interactions  
    per event for different values of \ptcorr. A fit of the exponential form $f=c_0+c_1\exp(c_2\cdot N_{\rm PV})$ is superposed
    where one degree of freedom is fixed via the constraint $f(0)=1$, i.e. $c_1=(1-c_0)$.}
    \label{fig:FractionOfQCDlikePUjets}
  \end{center}
\end{figure}

The contribution of particles from any vertex to a given \pujet{} can be measured using the jet vertex fraction (\JVF). It is defined as 
\begin{equation}
    \JVF(V_i) = \dfrac{\sum_{k = 1}^{\Npart(V_{i})} p_{\mathrm{T},k}}{\sum_{i=1}^{\Ncoll}\sum_{k=0}^{\Npart(V_{i})} p_{\mathrm{T},k}} = \dfrac{1}{\pT} \sum_{k = 1}^{\Npart(V_{i})} p_{\mathrm{T},k} ,
\label{eq:JVF}
\end{equation} 
where $\Npart(V_{i})$ is the number of particles from a given vertex $V_{i}$, and \Ncoll{} is the number of collision vertices contributing particles to the jet. \JVF{} is calculated for each of these vertices. 
Note that \pT\ corresponds to the uncorrected jet transverse momentum and consequently, the 
value of each component of $\JVF(V_i)$ depends on $\mu$. 

\subsection{Evaluation of the \pu{} jet nature}

It follows from the definition of \Rpt{} that \pujet s with values of \Rpt{} close to unity are matched to a truth jet with $\pT \approx \ptcorr$ of the \pujet{} itself. Consequently, there is a single \MB{} interaction which predominantly contributes to the jet. On the other hand, jets with a small value of \Rpt{} are mostly stochastic, as no single minimum-bias collision contributes in a dominant way to the \pujet. We characterize jets as stochastic if \Rpt{} is smaller than 0.8. This threshold value is arbitrary and the fraction of QCD-like and stochastic jets depends on the exact choice. The conclusion of our study holds for a broad range of cut values.

The fractions of QCD-like and stochastic \pujet s change as a function of pileup jet \pt{} and $\mu$. This can be seen in Fig.~\ref{fig:FractionOfQCDlikePUjets}, where \qcdjet-like
samples are defined by $\Rpt > 0.8$ for each \pu{} level. The fraction of these jets at a given \ptcorr{} decreases exponentially with $\mu$.  The exponential decrease is slower for larger the \ptcorr. 
At a \pu{} activity of $\mu=100$, the fraction of \pujet{}s that are QCD-like is about 40\% (20\%) for $\ptcorr > 40\GeV$ ($20 < \ptcorr <30\GeV$). At  $\mu=150$, these numbers
decrease to about 25\% and 15\%, respectively.

\subsection{\PU{} jet multiplicity}
\begin{figure*}[htbp!]\centering
\subfigure[]{\includegraphics[width=0.46\textwidth]{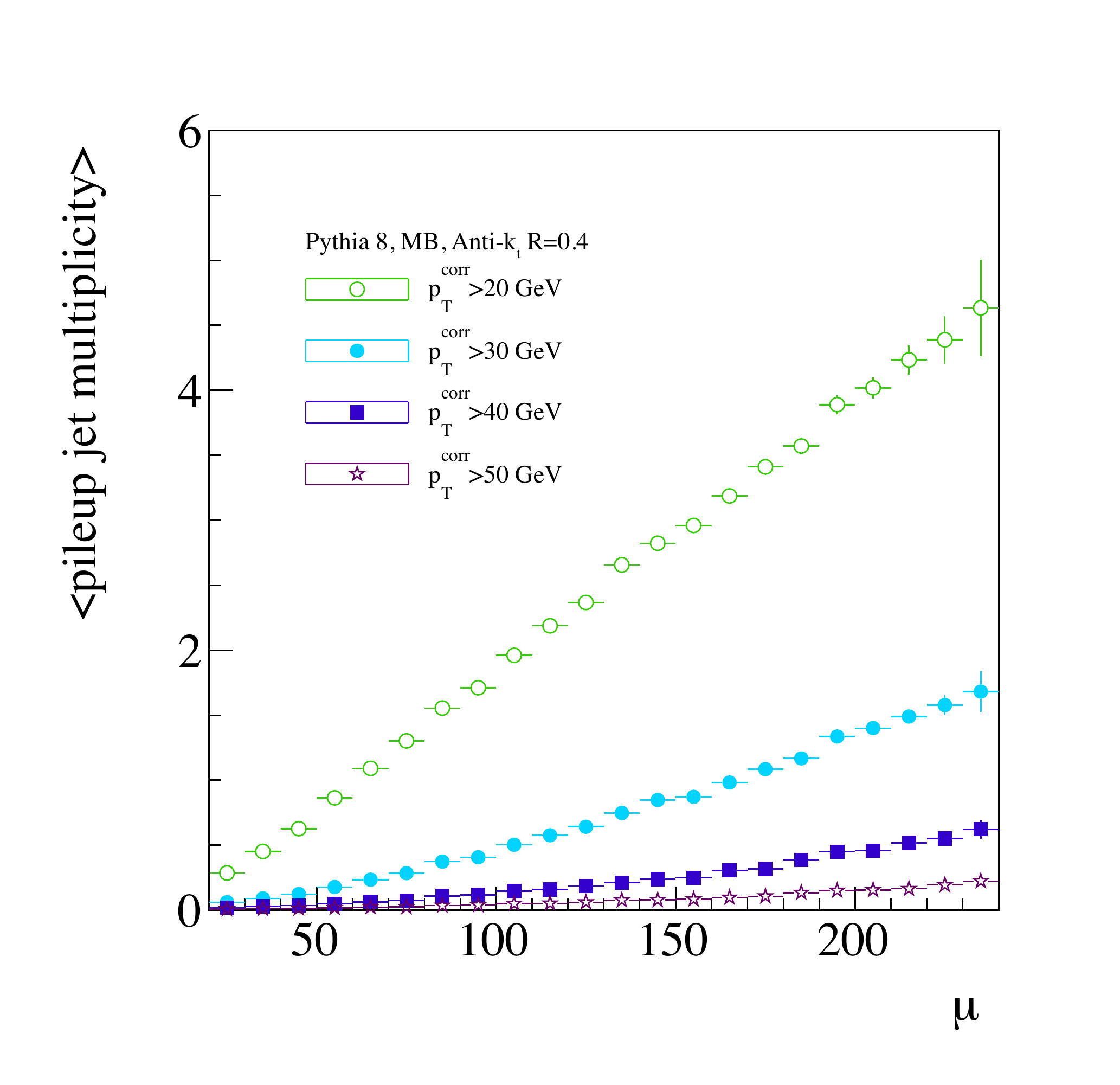}\label{fig:pujetmult:0}}
\subfigure[]{\includegraphics[width=0.46\textwidth]{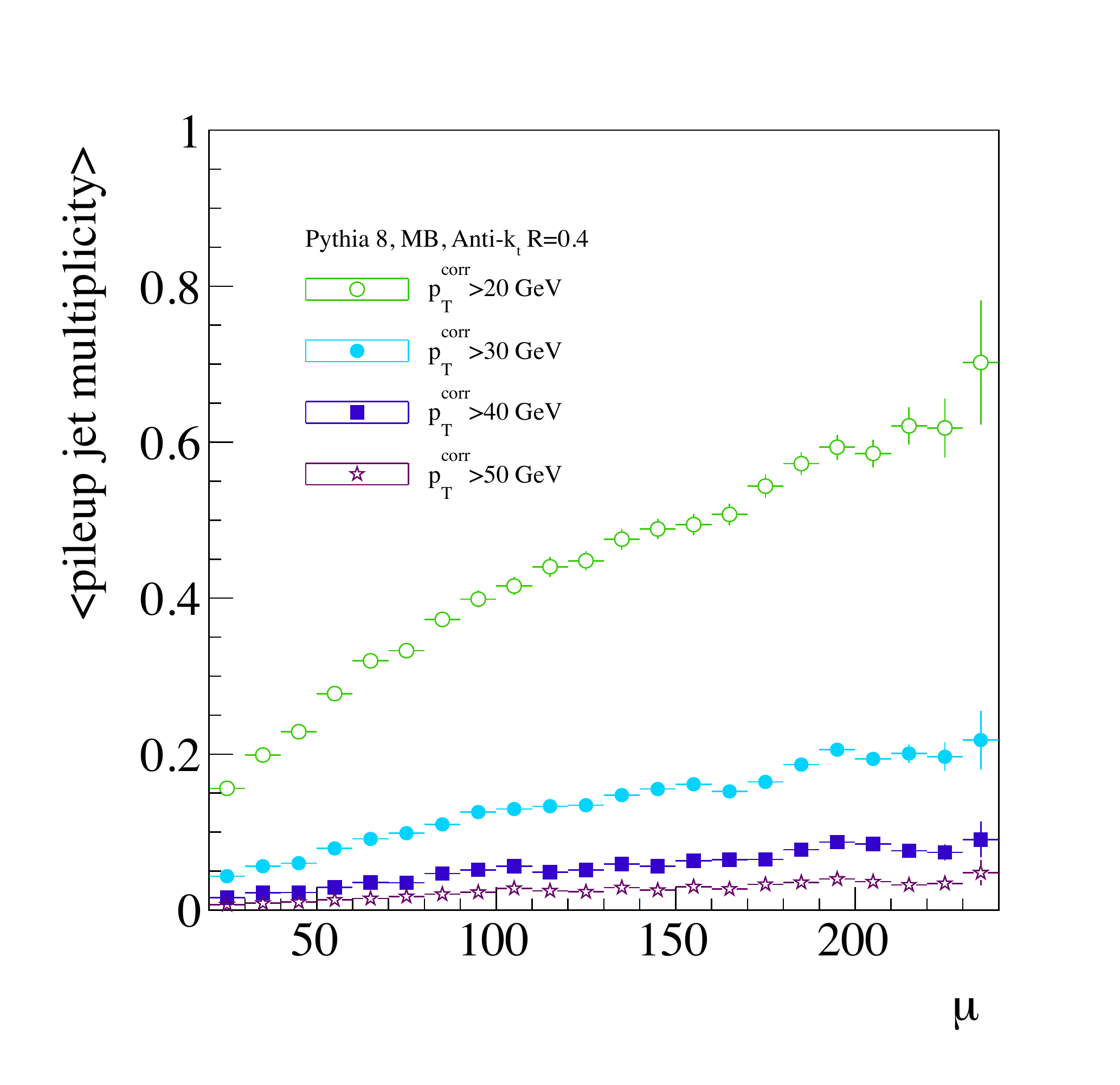}\label{fig:pujetmult:1}}
    \caption[]{The mean number of pileup jets per event inclusively \subref{fig:pujetmult:0} and for QCD-like pileup jets with $\Rpt > 0.8$
    \subref{fig:pujetmult:1}, as a function of $\mu$ and \ptcorr. }
    \label{fig:pujetmult}
\end{figure*}

The mean number of pileup jets per event, as a function of jet \ptcorr~and 
$N_{\rm PV}$, is indicative of the efficiency of the jet area based method to suppress jets generated by \pu. It is shown in Fig.~\ref{fig:pujetmult} for the inclusive \pujet s and separately for the subsample of QCD-like \pujet s satisfying $\Rpt  > 0.8$. It is observed that the average inclusive number $\langle N\rangle$ of low ($\ptcorr \simeq 20\GeV$) \pujet s per event increases rather linearly with $\mu$, i.e. $\partial \langle N \rangle/\partial \mu \approx \mathrm{const}$. For higher \pujet{} \pT, $\partial \langle N \rangle/\partial \mu$ is significantly smaller, and displays an increase with increasing $\mu$.

The sub-sample of QCD-like jets in the inclusive \pujet{} sample shows a different behavior, as indicated in the righmost panel of Fig.~\ref{fig:pujetmult}. In this case $\partial \langle N \rangle/\partial \mu$ decreases with increasing $\mu$ in all considered bins of \ptcorr. This contradicts the immediate expectation of an increase following the inclusive sample, but can be understood from the fact that with increasing $\mu$ the likelihood of QCD-like jets to overlap with (stochastic) jets increases as well. The resulting (merged) \pujet s no longer display features consistent with \qcdjet s (e.g., loss of single energy core), and thus fail the $\Rpt > 0.8$ selection.

The \pujet{} multiplicity shown in Fig.~\ref{fig:pujetmult} is evaluated as a function of the \pu{} corrected transverse momentum of the jet (\ptcorr). This means that after the correction approximately two \pujet{}s with $\ptcorr > \pTmin = 20$~\GeV{} can be expected for $\axing \simeq 100$. This number decreases rapidly with increasing \pTmin. The mean number of \qcdjet s is small, at about 0.4 at $\axing = 100$, for $\pTmin = 20$~\GeV.

\subsection{Jet grooming configurations} \label{sec:grooming}

Three jet grooming techniques are used by the LHC experiments:
\begin{description}
\item{\textbf{Jet trimming}} Trimming is described in detail in Ref.~\cite{Krohn:2009th}. In this approach the constituents of the large \antikt{} jet formed with $R = 1.0$ are re-clustered into smaller jets with $R_{\mathrm{trim}} = 0.2$, using the \antikt{} algorithm again. The resulting sub-jets are only accepted if their transverse momentum is larger than a fraction $f$ (here $f = 0.03$) of a hard scale, which was chosen to be the \pT{} of the large jet. The surviving sub-jets are recombined into a groomed jet.
\item{\textbf{Jet filtering}} Filtering was introduced in the context
  of a study to enhance the signal from the Higgs boson decaying into
  two bottom-quarks, see Ref.~\cite{Butterworth:2008iy}. In its
  simplified configuration without mass-drop criterion~\cite{Cacciari:2008gd} 
  applied in this study it works
  similar to trimming, except that in this case the sub-jets are found
  with the Cambridge-Aachen algorithm
  \cite{Wobisch:1998wt,Wobisch:2000dk} with $R_{\mathrm{filt}} = 0.3$,
  and only the three hardest sub-jets are retained. The groomed jet is
  then constructed from these three sub-jets.   
\item{\textbf{Jet pruning}} Pruning was introduced in
  Ref.~\cite{Ellis:2009su}. Contrary to filtering and trimming, it
  is applied during the formation of the jet, rather than based on
  the recombination of sub-jets. It dynamically suppresses small
  and larger distance contributions to jet using two parameters,
  $Z_{\mathrm{cut}}$ for the momentum based suppression, and
  $D_{cut} = D_{\mathrm{cut,fact}} \times 2 m/\pT$ (here $m$ and
  \pT{} are the transverse momentum and mass of the original jet)
  for the distance based. Pruning vetoes
  recombinations between two objects $i$ and $j$ for which the
  geometrical distance between $i$ and $j$ is more than $D_{\rm cut}$
  and the $p_T$ of one of the objects is less than $Z_{\rm cut}
  \times p_T^{i+j}$, where $p_T^{i+j}$ is the combined transverse
  momentum of $i$ and $j$. In this case, only the hardest of the two
  objects is kept. Typical values for the parameters are: 
  $Z_{\mathrm{cut}} = 0.1$ and
  $D_{\mathrm{cut,fact}} = 0.5$.
\end{description} 

In this study, trimming and filtering are applied to the original \antikt{} jets with size $R = 1.0$. 
%
%
We study the interplay between jet grooming and area-based pile-up 
correction. The subtraction is applied directly on the 4-momentum 
of the jet using:
\begin{equation}\label{eq:pusub}
  p^\mu_\text{jet,sub} = p^\mu_\text{jet}  
  - [\rho A^x_\text{jet}, \,\rho A^y_\text{jet}, \,
  (\rho+\rho_m) A^z_\text{jet}, \,(\rho+\rho_m) A^E_\text{jet}]\,,
\end{equation}
with
\begin{equation}
  \rho = \mathop{\text{median}}_\text{patches}
  \left\{\frac{p_{t,\text{patch}}}{A_\text{patch}}\right\}
  \,,\quad
  \rho_m = \mathop{\text{median}}_\text{patches}
  \left\{\frac{m_{\delta,\text{patch}}}{A_\text{patch}}\right\},
\end{equation}
$m_{\delta,\text{patch}} = \sum_{i\in\text{patch}}
\big(\sqrt{\smash[b]{m^2_{i} + p_{t,i}^2}} - p_{ti}\big)$, and $A^\mu$
is the active area of the jet as defined in
Ref.~\cite{Cacciari:2008gn} and computed by FastJet. The $\rho$ term,
mentioned above is the standard correction typically correcting the
transverse momentum of the jet. The $\rho_m$ term corrects for
contamination to the total jet mass due to the PU particle. 
When applying this subtraction procedure, we
discard jets with negative transverse momentum or (squared) mass of the jet.

\begin{figure*} [p!]
\begin{center}
\subfigure[raw, ungroomed jets]{\includegraphics[width=0.4\textwidth]{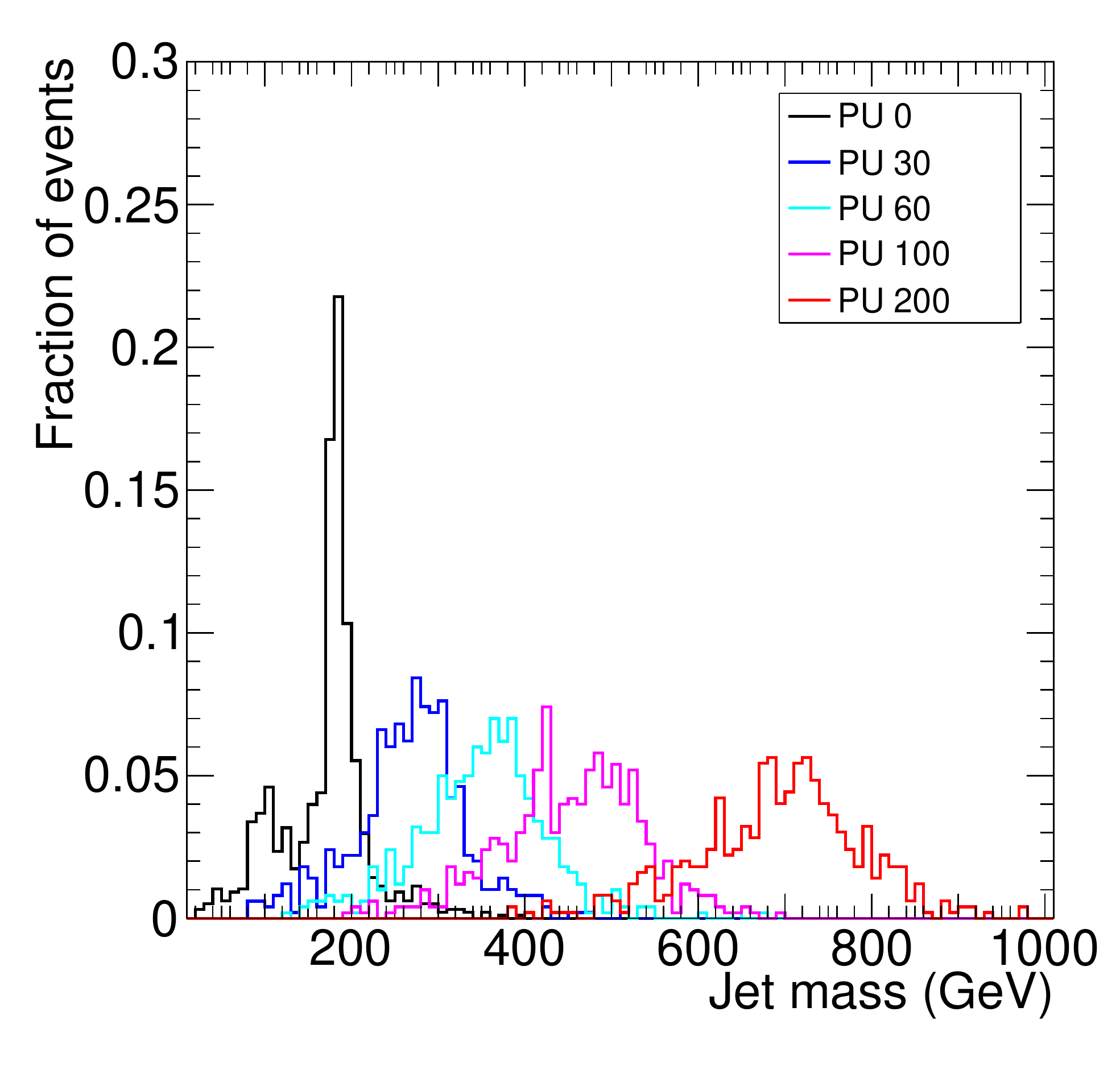}}
\subfigure[raw jets with pile-up subtraction]{\includegraphics[width=0.4\textwidth]{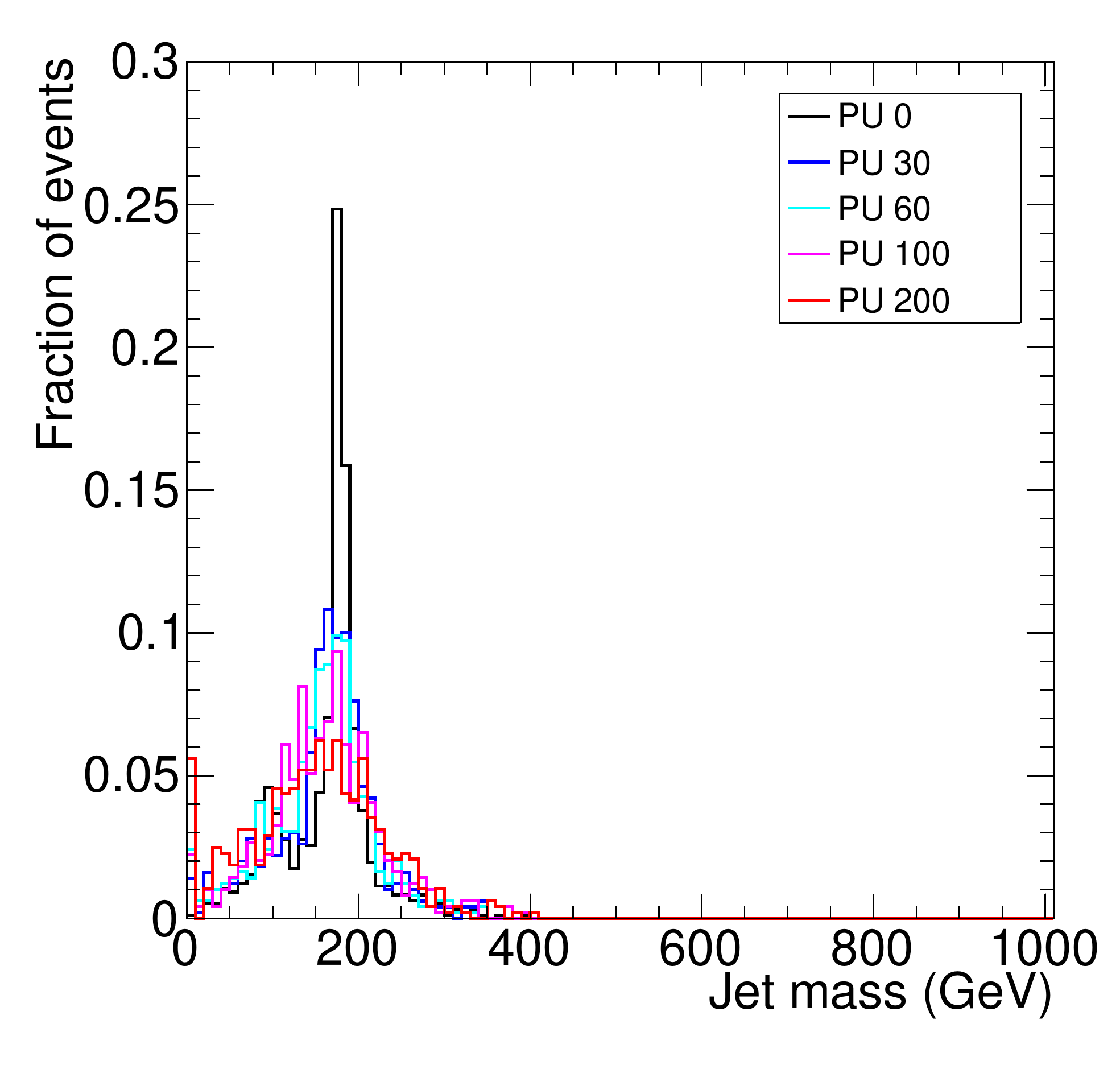}}
\subfigure[trimmed jets]{\includegraphics[width=0.4\textwidth]{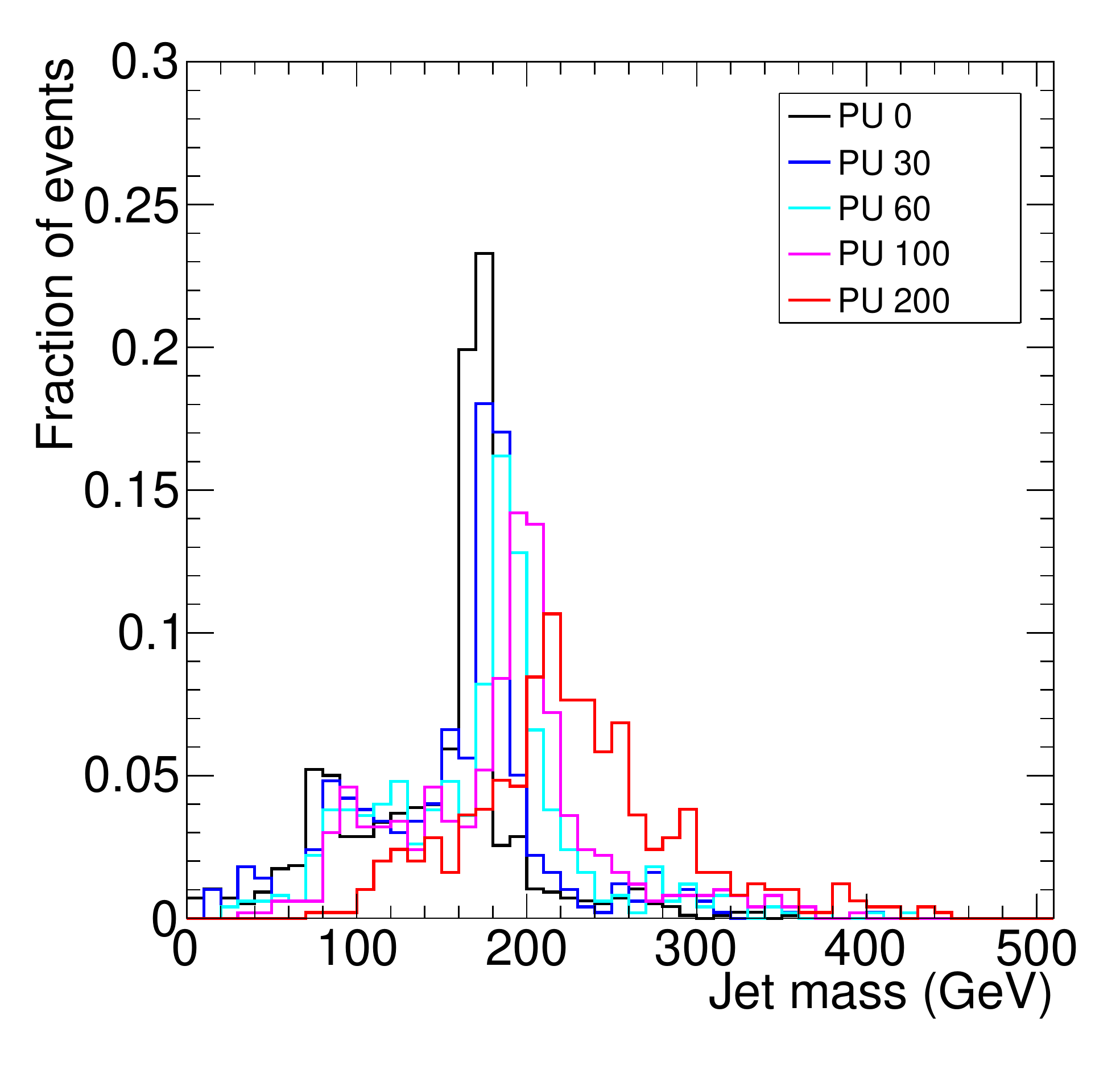}} 
\subfigure[trimmed jet with pile-up subtraction]{\includegraphics[width=0.4\textwidth]{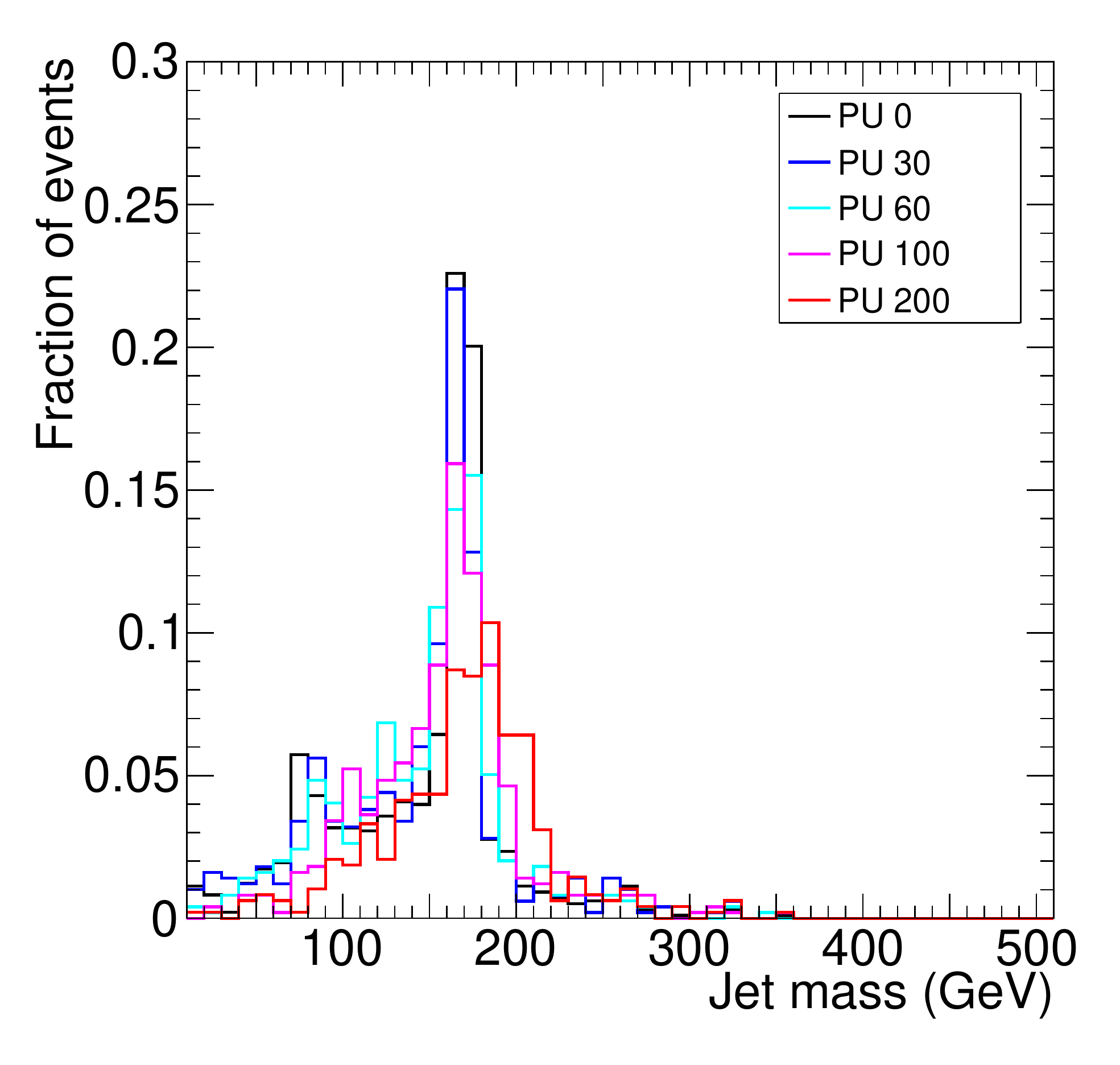}}
\subfigure[filtered jets]{\includegraphics[width=0.4\textwidth]{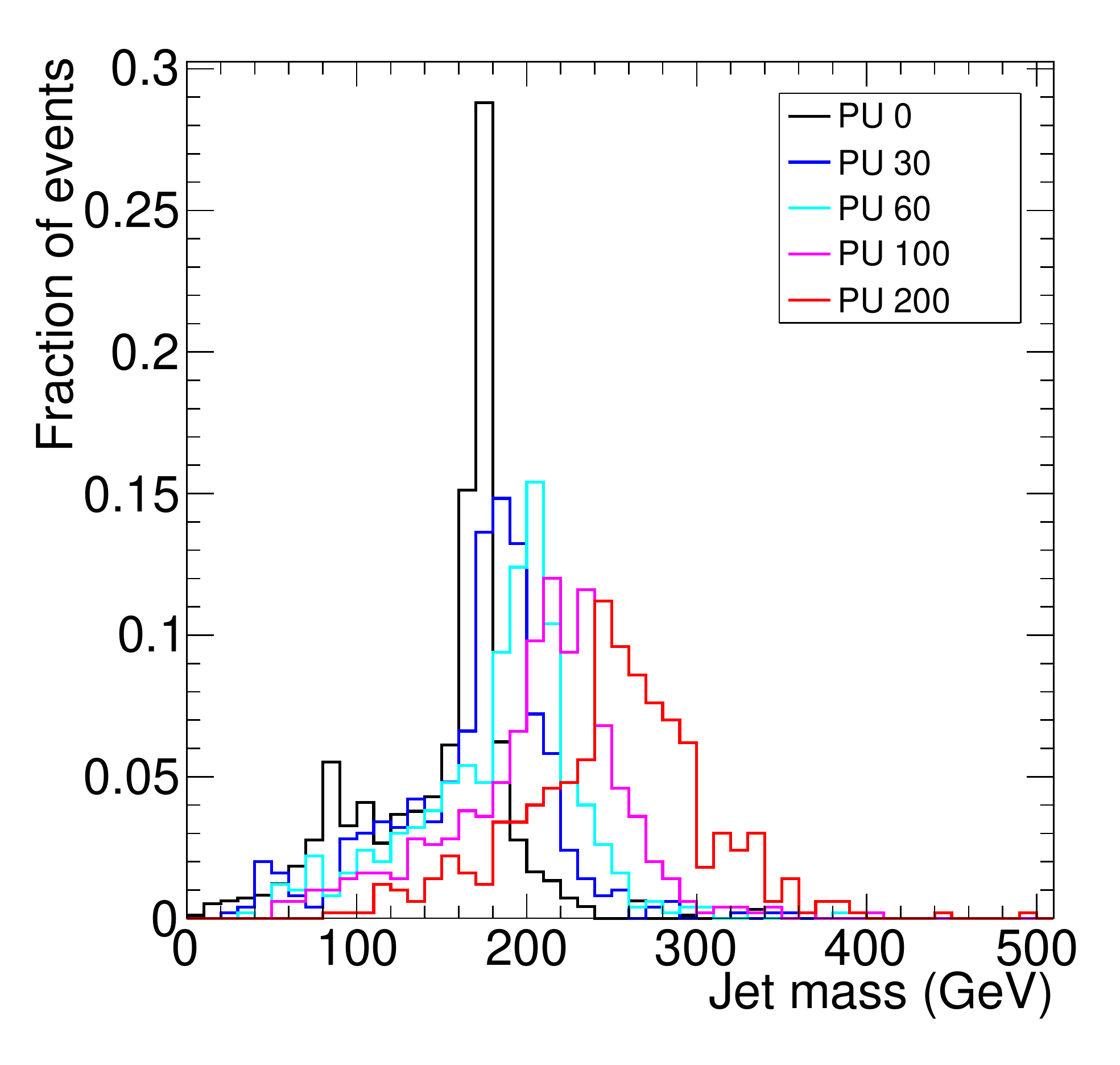}}
\subfigure[filtered jet with pile-up subtraction]{\includegraphics[width=0.4\textwidth]{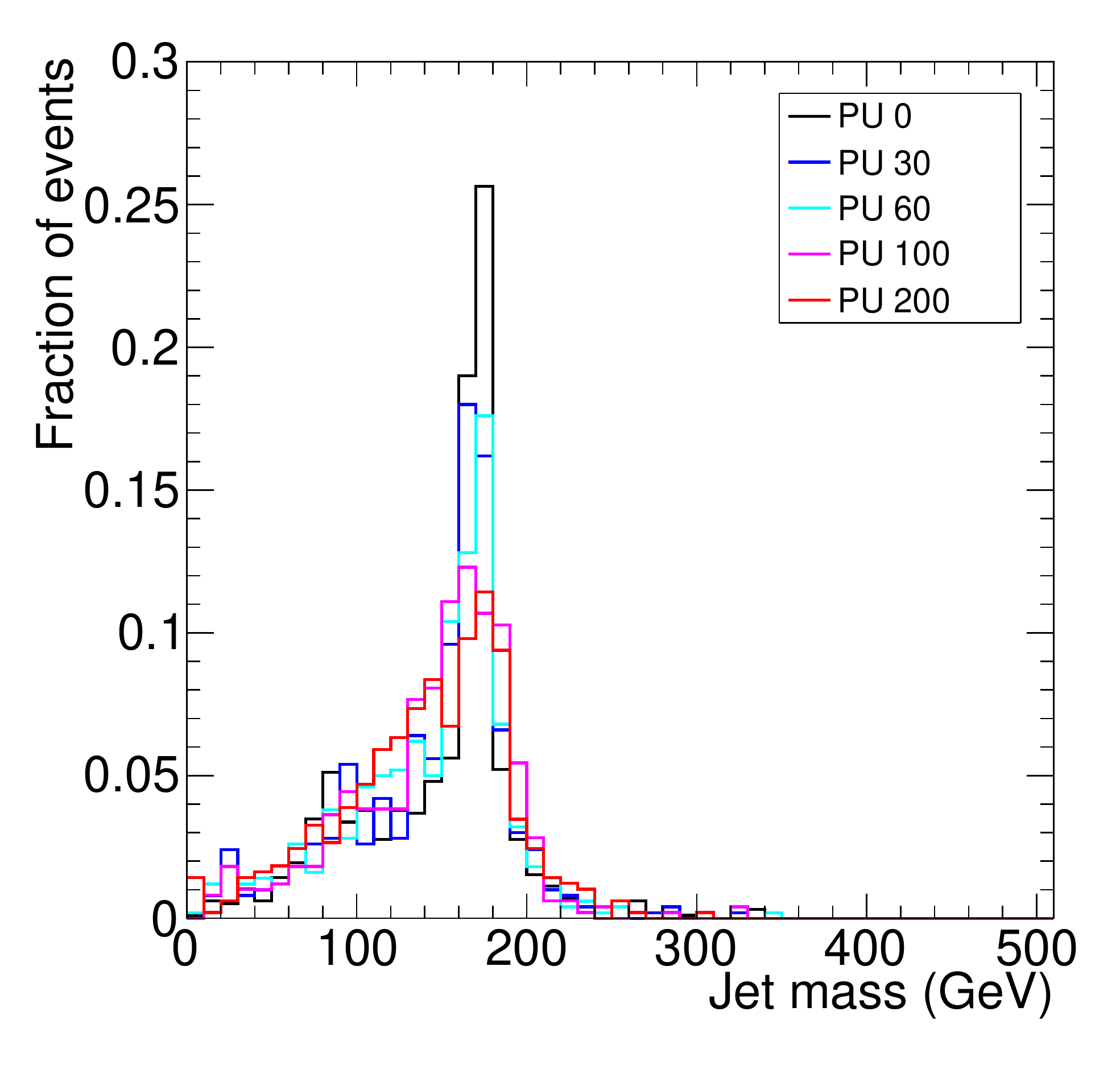}} 
\end{center}
\caption[]{The impact of \pu{} on the jet mass distribution. Top row: the raw jet mass distribution for jets reconstructed with the \antikt{} algorithm and $R = 1.0$ in $\Zprime \to \ttbar$ final states with $m_{Z'} =$ 1.5~\tev{}, in the presence of pile-up with $\axing = 30$, 60, 100, and 200, before and after pile-up subtraction. The second and third row show the same result after trimming (middle row) and filtering (lower row).}
\label{fig:mass_spectra}
\end{figure*}

The estimation of $\rho$ and $\rho_m$ is performed with \FJ\ 
using\footnote{Ghosts
  are placed up to $y_{\rm max}=3$ and explicit ghosts are enabled.}
$k_t$ jets with $R=0.4$. Corrections for the rapidity dependence of
the pileup density $\rho$ are applied using a rapidity rescaling.

When we apply this background subtraction together with trimming or
filtering, the subtraction is performed directly on the subjets,
before deciding which subjets should be kept, so as to limit the
potential effects of pileup on which subjets are to be kept.

\subsection{Jet substructure performance}
\label{sec:results}

%

The various methods and configurations discussed in the previous
section are applied to the jets reconstructed with the \antikt{}
algorithm with $R = 1.0$ in the $\Zprime \to \ttbar$ final state in
the presence of pile-up. For the studies presented in this report we require
jet \pT{} before grooming and pileup subtraction to be greater than 100
GeV and consider the two hardest \pT jets in the event. We further
require that the rapidity difference between the two jets $|y_1 -
y_2|$ is less than one.
The immediate expectation for the reconstructed jet mass $m$ is the
top mass, i.e. $m \approx 175$~\GeV, and no residual dependence on the
pile-up activity given by \axing, after the pile-up
subtraction. The two plots in the upper row of 
Figure \ref{fig:mass_spectra} show the distributions of
the reconstructed jet masses without any grooming
 and with the pile-up subtraction discussed in Section \ref{sec:grooming} 
applied. The effect of
pile-up on the mass scale and resolution is clearly visible.
Applying only the pile-up subtraction, without changing
the composition of the jets, already improves the mass reconstruction
significantly. All \axing{} dependence is removed from the jet mass spectrum,
as shown in Fig.~\ref{fig:mass_spectra}. In
particular, the position of the mass peak is recovered. With
increasing pileup, the mass peak gets more and more smeared, an effect
due to the fact that the pileup is not perfectly uniform. These
point-to-point fluctuations in an event lead to a smearing $\pm
\sigma\sqrt{A}$ in~\eqref{eq:pusub}. For very large pileup, this
smearing extends all the way to $m=0$ as seen in
Fig.~\ref{fig:mass_spectra}.

The effect of the other grooming techniques on the reconstructed jet mass distributions is summarized in Fig.~\ref{fig:mass_spectra}, with and without the pile-up subtraction applied first. The spectra show that both trimming and filtering can improve the mass reconstruction. The application of the pile-up subtraction in addition to trimming or filtering further improves the mass reconstruction performance.

\begin{figure*} [htbx!]
\begin{center}
\includegraphics[width=0.45\textwidth]{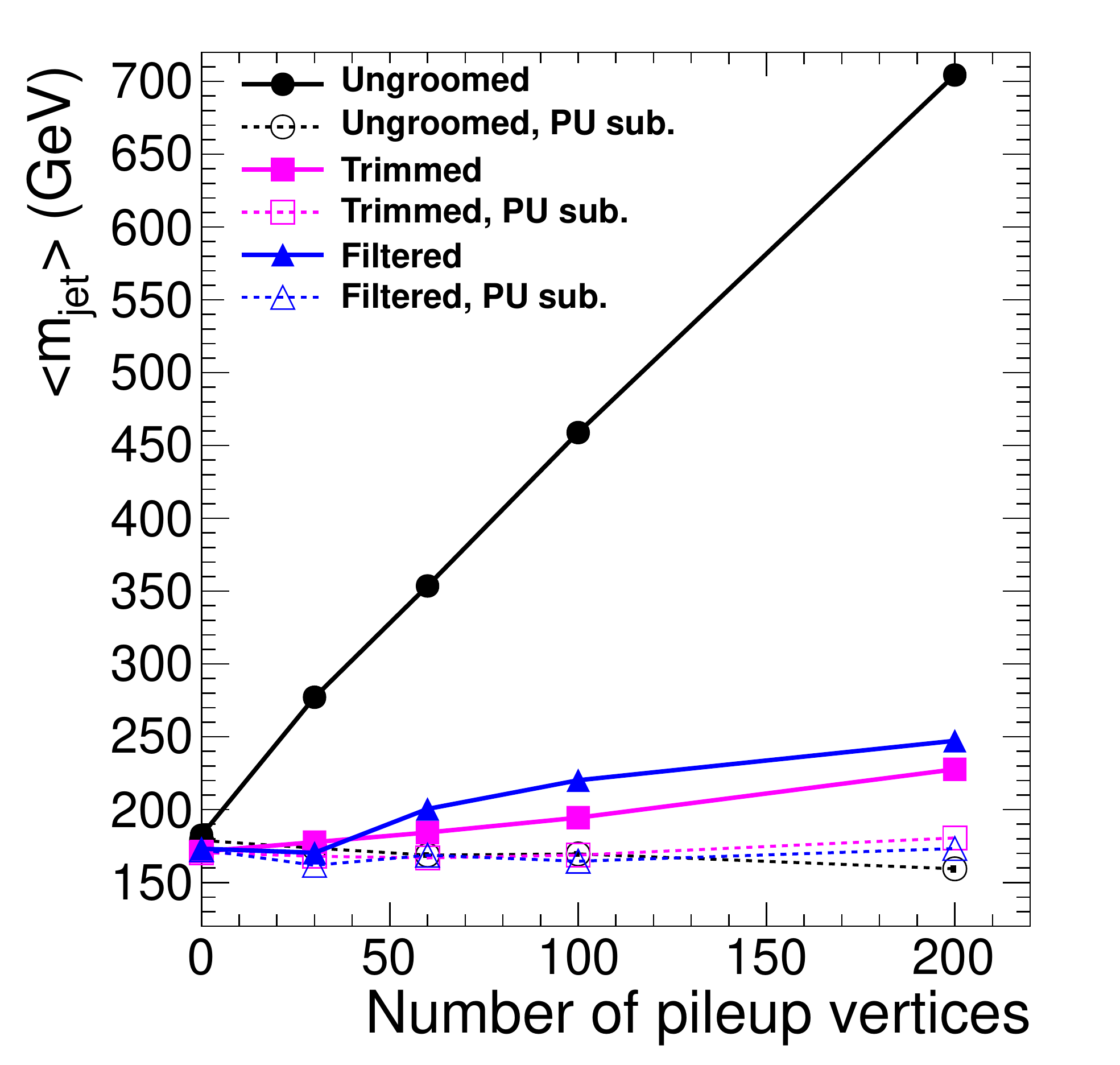}
\includegraphics[width=0.45\textwidth]{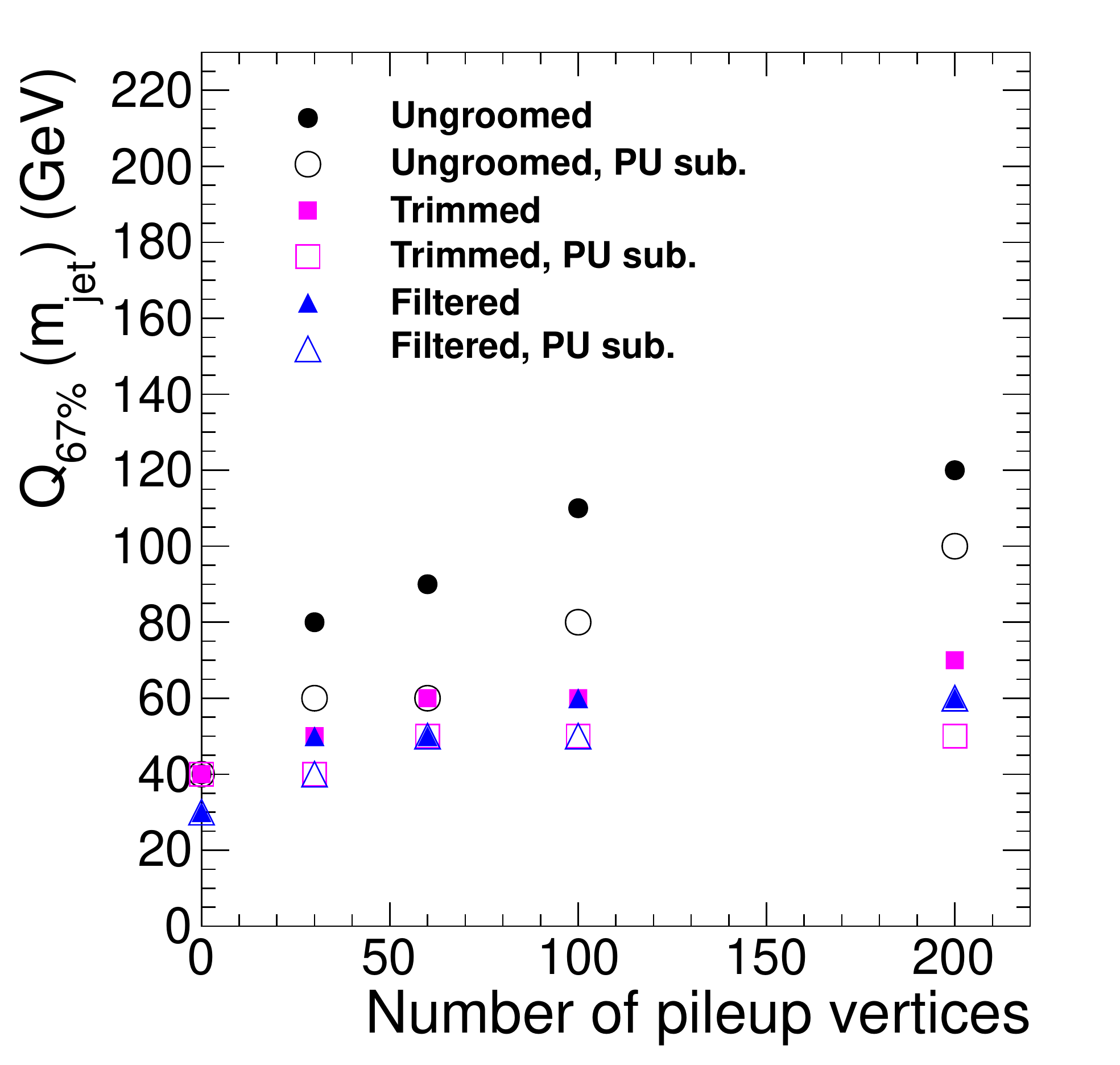}
\caption[]{Average (leftmost figure) and RMS (rightmost figure) of the reconstructed jet mass distribution in $\Zprime \to \ttbar$ final states, as a function of the pile-up activity \axing, for various jet grooming techniques.}
\label{fig:mass_summary}
\end{center}
\end{figure*}

The findings from the spectra in Figs.~\ref{fig:mass_spectra} are quantitatively summarized in Fig.~\ref{fig:mass_summary} for the mass scale and resolution. Here the resolution is measured in terms of the mass range in which $67\%$ of all jet masses can be found ($Q_{67\%}(m_{\mathrm{jet}})$ quantile).  
Maintaining the jet mass scale around the expectation value of 175~\GeV{} works well for trimming and filtering with and without pile-up subtraction, see Fig.~\ref{fig:mass_summary}. The same figure indicates that for very high pile-up ($\axing = 100 - 200$), the jet mass after trimming and filtering without pile-up subtraction shows increasing sensitivity to the pile-up. The additional pile-up subtraction tends to restore the mass scale with better quality.

Both trimming and filtering improve the mass resolution to different degrees, but in any case better than pile-up subtraction alone, as expected. Applying the additional pile-up subtraction to trimming yields the least sensitivity to the pile-up activity in terms of mass resolution and scale. 
%


These effects can be explained as follows. As discussed earlier,
pileup has mainly two effects on the jet: a constant shift
proportional to $\rho A$ and a smearing effect proportional to $\sigma
\sqrt{A}$, with $\sigma$ a measure of the fluctuations of the pileup
within an event. In that language, the subtraction corrects for the
shift leaving the smearing term untouched. Grooming, to the contrary,
since it selects only part of the subjets, acts as if it was reducing
the area of the jet\footnote{Note that grooming techniques do more
  than reducing the catchment area of a jet. Noticeably, the selection
  of the hardest subjets introduces a bias towards including upwards
  fluctuations of the background. This positive bias is balanced by a
  negative one related to the perturbative radiation discarded by the
  grooming. These effects go beyond the generic features explained
  here.}. This reduces both the shift and the dispersion. Combining
grooming with subtraction thus allows to correct for the shift
leftover by grooming and reduce the smearing effects at the same
time. All these effects are observed in Figures~\ref{fig:mass_summary}.


\subsection{Concluding remarks}\label{subsec:concl}

The source of jets produced in minimum bias collisions in the presence of \pu{} is analyzed using a technique relating the single collision contribution in the jet to its transverse momentum after \pu{} correction in particle level Monte Carlo. The rate of \pujet s surviving after application of the jet area based \pu{} subtraction is about two with $\ptcorr > 20$~\GeV{} and within $|y|<2$, at a \pu{} activity of $\axing = 100$. It rises about linearly with increasing pile-up for this particular selection. Higher \pT{} jets occur at a much reduced rate, but with a steeper than linear rise with increasing $\mu$. 

The rate of QCD-like jets is significantly smaller, and shows a less-than-linear increase with increasing $\mu$ even for $\pTmin = 20$~\GeV. This can be understood as a sign of increased merging between QCD-like jets and stochastic jets. The merged jets are less likely to display features characteristic for QCD-like jets, and therefore fail the selection. 

The fraction of QCD-like jets with a core of energy arising from a single \pp{} interaction of at least $0.8 \ptcorr$ is found to decrease rapidly with
increasing $\mu$. At $\mu=50$ about 60\% of the \pu{} jets with $\ptcorr > 50\GeV$ are found to be QCD-like, whereas at $\mu=200$ this number is decreased to 
about 20\%.

A brief Monte Carlo study of the effect of jet grooming techniques on
the jet mass reconstruction in $\Zprime \to \ttbar$ final states has
been conducted. Jet trimming and filtering are used by themselves, or
in combination with the pile-up subtraction using the four-vector
area, to reconstruct the single jet mass and evaluate the stability of
the mass scale and resolution at pile-up levels of 30, 60, 100, and
200 extra \pp{} collisions, in addition to the signal event. It is
found that for this particular final state trimming and filtering work
well for maintaining the mass scale and resolution, provided they are
applied together with pile-up subtraction so as to benefit both from
the average shift correction from subtraction and noise reduction from
the grooming. 
%

The studies presented here are performed with Monte Carlo simulated signal and pile-up (minimum bias) interactions. No considerations have been given to detector sensitivities and other effects deteriorating the stable particle level kinematics and flows exploited here. With this respect the conclusions of this study are limited and can be considered optimistic until shown otherwise.     
Note also that comparing the performance of filtering and trimming
would require varying their parameters and that this goes beyond the
scope of this study.




\section{The potential of boosted top quarks}
\label{sec:top}
{\it Section prepared by the Working Group: 'Prospects for boosted top quarks',  A. Altheimer, J. Ferrando, \underline{J. Pilot}, S. Rappoccio, M. Villaplana, \underline{M. Vos}.
}

Among the many applications of the strategies for boosted objects
discussed in the literature 
(the bibliography of References~\cite{Abdesselam:2010pt,Altheimer:2012mn} is
a good starting point to navigate the extensive literature), the study of 
highly energetic top quarks forms 
the case that has been studied in greatest detail by the experiments. 
Several studies of the production of boosted top
quarks have set limits on new physics scenarios. The first sample 
of boosted top quarks has also been used to understand the 
modelling of the parton shower and the detector response. In this section
we present a summary of achievements so far, discuss how existing
analyses could benefit from an improved understanding of jet substructure,
and explore possible directions for future work.

\subsection{Boosted top quark production}

\begin{table*}[htbp!]
\centering
\caption{The top pair production rate at past, present and future colliders, 
calculated with the MCFM code~\cite{Campbell:2010ff}. The inclusive 
production rate 
is given in the first row. The expected number of events with boosted 
top quarks ($M_{t\bar{t}} > $ 1~\tev) and highly boosted top quarks 
($M_{t \bar{t}} > $ 2~\tev) is given in the second and third row, respectively.}
\begin{tabular}{|l|c|c|c|c|} \hline
Collider \& phase & Tevatron run II & LHC 2012 & LHC phase II & HE-LHC \\
process \& energy,  &  $ p \bar{p}$ at $\sqrt{s} = $ 1.96~\tev & $pp $ at  $\sqrt{s} = $8~\tev &  $pp $ at  $\sqrt{s} = $13~\tev & $pp $ at  $\sqrt{s} = $33~\tev \\
integrated luminosity      & $\mathcal{L} =$ 10~\ifb{}  & $\mathcal{L} =$ 20~\ifb{} & $\mathcal{L} =$ 300~\ifb{} & $\mathcal{L} =$ 300~\ifb{}  \\ \hline
Inclusive $t \bar{t}$ production & 6 $\times $ 10$^{4}$ & 4 $\times$ 10$^6$ & 2 $\times $ 10$^8$ & 1.4 $\times $ 10$^9$  \\ \hline
Boosted production & 23 &  6 $\times $ 10$^{4}$ &  5.2 $\times$ 10$^6$ &  7.1 $\times$ 10$^7$ \\ \hline
Highly boosted & 0 &  500 &  1.1 $\times$ 10$^5$ & 3.9 $\times$ 10$^6$ \\ \hline
\end{tabular}
\label{tab:boostedtoprates}
\end{table*}

The top quark decay topology observed in the detector depends strongly on the
kinematic regime.  The decays products of top quarks produced nearly at rest 
($p_T < 200$ GeV/$c$) are well-separated, leading to experimental signatures
such as isolated leptons and a relatively large number of clearly resolved 
jets. With increasing transverse momentum, the decay products 
of the top quark will become collimated and possibly reconstructed in the 
same final state object. For intermediate boosts (200 $< p_T < $400~\gev{}), 
the daughters of the $W$ boson from a fully-hadronic top decay will be close 
enough to be clustered into the same jet.  At this point, the use of jet 
substructure techniques becomes important to efficiently identify these decay 
signatures. At even larger $p_T$ top quarks become truly boosted objects: 
all decay products of the top will be 
strongly collinear, with the $\Delta R \sim 2 m_{\mathrm{top}} / p_T$.  
Hadronic top quarks can be reconstructed in a single 
jet, and top quarks with leptonic decays generally contain non-isolated 
leptons due to the overlap with the $b$-quark jet.

Table~\ref{tab:boostedtoprates} presents the expected numbers of 
boosted top quark pairs according to the Standard Model at past, present and 
future colliders. The numbers show clearly how the study of boosted top quarks 
becomes viable only with the start of the LHC. The first phase of operation
yields a sample of several tens of thousands of boosted top quark pairs.
The next-to-last column indicates the size of the 
sample expected in a 13 or 14~\tev{} run of the LHC, that is to start 
by the end of 2014. The increase in the centre-of-mass 
energy and the larger integrated luminosity each bring an increase 
of an order of magnitude in the production of boosted top quarks.

We expect, therefore, that boosted topologies will gain considerable importance 
as the LHC program develops. To exploit the LHC data to their full potential it 
is  critical that existing experimental strategies are adapted to this
challenging kinematical regime. Before we turn to the results of 
analyses of boosted object production, we discuss a number of new tools
that were developed to identify and reconstruct boosted top quarks efficiently.

\subsection{Top Tagging.}  

Excellent reviews of top tagging algorithms exist~\cite{Plehn:2011tg}. 
Previous BOOST reports have compared their 
performance for simulated events (at the particle level). In this 
Section we present a very brief review for completeness. 

The Johns Hopkins (JHU) tagger \cite{Kaplan:2008ie} identifies 
substructure by reversing through the iterative clustering process 
used to form jets.  Subjets are found using several criteria -- the 
ratio of their individual $p_T$ to the original jet $p_T$ must be above 
a given threshold, and the subjets must be spatially separated from each 
other to give a valid decomposition.  In this way, a jet can be 
deconstructed into up to four subjets, and jets with three or more 
subjets are analyzed further, requiring the invariant mass of the 
identified subjets to be in the range $[145, 205]$~\gev{}, and two 
of the subjets to be consistent with $m_W$, in the range $[65,95]$~\gev.  
There is an additional cut on the $W$ boson helicity angle, 
$\cos \theta_h < 0.7$. 

The variant of the JHU tagger used by CMS\cite{CMS-PAS-JME-09-001} uses 
a similar jet decomposition, with slight differences in the selections of 
top quark and $W$ boson masses from the subjets.  Additionally, the CMS 
top tagger does not apply the $W$ boson helicity angle requirement, but
instead selects jets with the
minimum pairwise mass of the subjets larger than 50~\gev.
The JHU and CMS top tagging algorithms have been developed with jet 
distance parameters up to $R = 0.8$, and therefore are only efficient 
for top quarks with $p_T$ above approximately 400 GeV/$c$.  

The HEP top tagger \cite{Plehn:2010st}, is designed to use jets with 
distance parameter $R=1.5$, thereby extending the reach of the tagging 
algorithm to lower jet $p_T$ values.  The algorithm uses a mass drop 
criterion to identify substructure within the jet, but also uses a 
filtering algorithm to remove soft and large-angle constituents from the 
individual subjets.  The three subjets with a combined mass closest to 
$m_t$ are then chosen for further consideration.  Cuts are then applied 
to masses of subjet combinations to ensure consistency with $m_W$ and 
$m_t$.  Specifically, for the three subjets sorted in order of subjet 
$p_T$, having masses $m_1, m_2, m_3$, the quantities $m_{23}/m_{123}$ 
and $\arctan m_{13}/m_{12}$ are computed.  Geometrical cuts can be 
applied in the phase space defined by these two quantities 
to select top jets and reject quark or gluon jets.  

The HEP top tagger obtains tagging efficiencies of up to 37\% for lower 
$p_T$ top quarks ($p_T > 200$ GeV/$c$), with an acceptable mistag rate.  
It has been used by the ATLAS \ttbar{} resonance search in the fully
hadronic channel~\cite{Aad:2012raa}, where no {\it resolved} analysis
has been performed. At high jet $p_T$, the efficiencies for the HEP Top 
Tagger and JHU Top Tagger selections are comparable.

Boosted top quarks were also studied using both $R=1.0$ anti-$k_{t}$ jets 
and jets identified by the HEPTopTagger~\cite{Plehn:2010st} algorithm 
as candidate ``top-jets.'' Kinematic and substructure distributions were 
compared between data and MC simulation and were found to be in agreement. 
Furthermore, the efficiency with which top quarks were identified as such 
was found to be significantly increased in both cases, and the HEPTopTagger 
was shown to reduce the backgrounds to such searches dramatically, even 
with a relatively relaxed transverse momentum selection.

Overall, the results from ATLAS suggests that, among the jet grooming 
configurations tested, the trimming algorithm exhibited an improved mass 
resolution and smaller dependence of jet kinematics and substructure 
observables on pile-up (such as 
$N$-subjettiness~\cite{Thaler:2010tr,Thaler:2011gf} and the 
$k_{t}$ splitting scales~\cite{Butterworth:2002tt}) compared to the 
pruning configurations examined. For boosted top quark studies, 
the anti-$k_{t}$ algorithm with a radius parameter of $R=1.0$ and 
trimming parameters $f_{\rm cut}=0.05$ and $R_{\rm sub}=0.3$ was found to 
be optimal, where a minimum $p_{T}$ requirement of 350 GeV is typical. It 
is important to note that only the $k_{t}$-pruning for $R=1.0$ jets was 
tested and that since the performance does depend somewhat on this 
parameter, further studies are necessary to optimize for other jet size. 
Lastly, Cambridge-Aachen jets with $R=1.2$ using the mass-drop filtering 
parameter $\mu_{\rm frac}=0.67$ were found to perform well for 
boosted two-pronged analyses such as $H\rightarrow b\overline{b}$ or 
searches involving boosted $W\rightarrow q\overline{q}$ decays.

A final algorithm that is currently being investigated is the $N$-subjettiness 
algorithm \cite{Thaler:2010tr} presented in Section~\ref{sec:mc}.

Several new techniques and ideas are emerging, that aim to improve 
boosted top identification and reconstruction.

One such technique is that of shower deconstruction \cite{Soper:2010xk}.  
This method aims to identify boosted hadronic top quarks by computing the 
probability for a top quark decay to produce the observed jet, including its 
distribution of constituents.  The probability for the same jet to have 
originated from a background process is also computed.  These probabilities 
are computed by summing over all possible shower formations resulting in the 
observed final state, accounting for different gluon splittings and 
radiations, among other processes. This is done both for the signal shower 
processes and background shower processes. A likelihood ratio is formed 
from the signal and background probabilities and used to discriminate 
boosted top quarks from generic QCD jets.  The process of evaluating all 
shower histories can be computationally intensive, so certain requirements 
are made on the number of constituents used in the method to make the 
problem tractable.  The results presented in Ref.~\cite{Soper:2012pb} 
show an improvement on the top taggers described previously. Specifically, 
the shower deconstruction method reduces the top mistag rate by a factor 
of 3.6 compared to the JHU top tagger, while maintaining the same signal 
acceptance.  This method is also applicable to the lower $p_T$ regime, 
and there improves upon the top mistag rate from the HEP top tagger by a 
factor of 2.6, again keeping identical signal efficiency.

\begin{figure*}[htpb!]
\begin{center}
\includegraphics[width=0.6\linewidth]{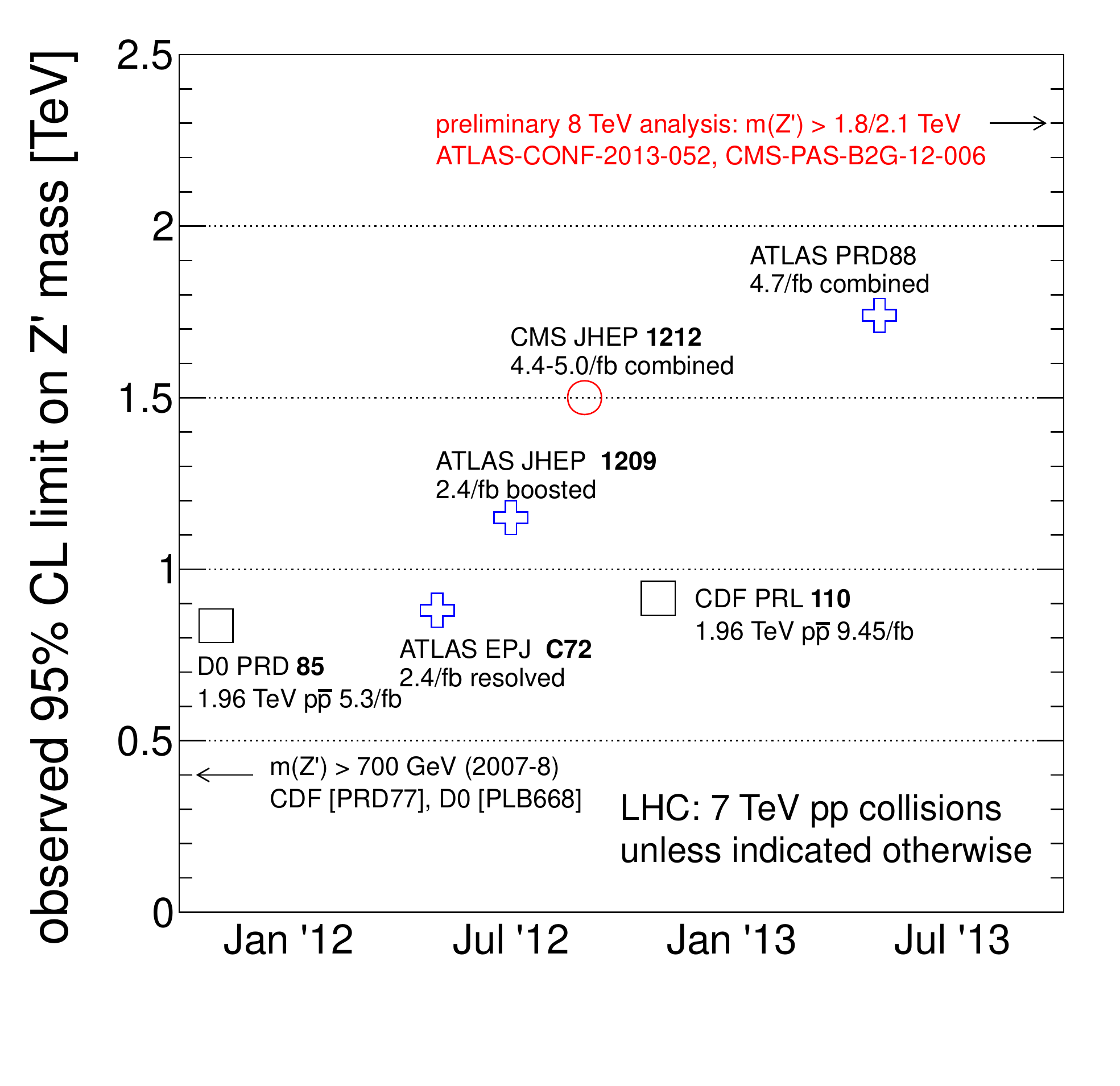}
\end{center}
\caption{Overview of evolution of the sensitivity of $t\bar{t}$ resonance searches in the first years of LHC operation. The sensitivity is presented in terms of the lower limit on the mass of a narrow $\Zprime$ boson. The production rate for this new state is given by a benchmark model that is common to all experiments (a leptophobic topcolor $\Zprime{}$ boson).}
\label{fig:summary_searches}
\end{figure*}

\begin{table*}[htpb!]
\begin{center}
\caption{Exclusion limits at 95\% confidence level for a narrow $\Zprime{}$ boson, as obtained in \ttbar{} resonance searches at the Tevatron and the first years of operation of the LHC.}
\vspace{2mm}
\begin{tabular}{lccccc}
\hline 
CDF and D0 References                        &  ~\cite{Aaltonen:2007ag} & ~\cite{Abazov:2008ny} &  ~\cite{Aaltonen:2011vi} & ~\cite{Abazov:2011gv} & ~\cite{Aaltonen:2012af}\\
Final state   \&                 & $l$+jets & $l$+jets & fully had. & $l$+jets & $l$+jets \\   
Reconstruction                   &  resolved & resolved & resolved & resolved & resolved \\  
$\sqrt{s}$ [\tev{}] & 1.96 &  1.96 &  1.96 & 1.96  & 1.96  \\
$\int{L} $ [~\ifb{}] & 1~\ifb{} & 1~\ifb{} &  4~\ifb{}  &  4~\ifb{} & 10~\ifb{} \\
\Zprime{} mass [TeV]  & $<$ 0.7 & $<$ 0.720 & $<$ 0.805 &   $ < $  0.835 & $ < $ 0.915  \\ \hline
ATLAS Reference                        &     ~\cite{Aad:2012wm}     &  ~\cite{Aad:2012dpa}  &  ~\cite{Aad:2012raa}  & ~\cite{Aad:2013nca}  &  ~\cite{ATLAS-CONF-2013-052} \\
Final state   \&                 & $l$+jets & $l$+jets & fully had. & $l$+jets & $l$+jets \\   
Reconstruction                   & resolved & boosted & boosted & combined & combined \\ 
$\sqrt{s}$ [\tev{}] &  7&  7 & 7  & 7 & 8 \\
$\int{L} $ [~\ifb{}] & 2.04~\ifb{} &   2.04~\ifb{} &  4.07~\ifb{}  &  4.07~\ifb{} & 14~\ifb{} \\
\Zprime{} mass [TeV]  & 0.5 $-$ 0.88 & 0.6 $-$ 1.15 & 0.7$-$1, 1.28$-$1.32  &   $ < $  1.74 & $ < $ 1.8 \\
$g_{KK} $ mass [TeV]  & 0.5 $-$ 1.13 & 0.6 $-$ 1.5  & 0.7 $-$ 1.62 & $<$ 2.07 & $ < $  2.0 \\ \hline 
CMS Reference                        &  ~\cite{Chatrchyan:2012ku}  &  ~\cite{Chatrchyan:2012cx}  & ~\cite{Chatrchyan:2012yca}  &  ~\cite{CMS-PAS-B2G-12-005} & ~\cite{CMS-PAS-B2G-12-006}\\
Final state   \&                 & fully hadronic  & $l$+jets & di-lepton & fully hadronic & $l$+jets\\   
Reconstruction                   & boosted         & combined &          & boosted & combined  \\ 
$\sqrt{s}$ [\tev{}] & 7 &  7 & 7  & 8 & 8 \\
$\int{L} $ [~\ifb{}] & 5.0~\ifb{} & 4.4$-$5.0~\ifb{} &   5.0~\ifb{} &  19.6~\ifb{}  &  19.6~\ifb{}  \\
\Zprime{} mass [TeV]  & 1.3 $-$ 1.5 & $ < $1.49 & $ < $1.3  & $ < $ 1.7 &   $ < $  2.1   \\
$g_{KK} $ mass [TeV]  &  1.4 $-$ 1.5  & $ < $1.82 & $ < $1.8  & $ < $ 1.8 & $ < $ 2.5   \\ \hline 
\end{tabular}
\label{tab:limits}
\end{center}
\end{table*} 

Another algorithm under development is the template 
overlap method~\cite{Backovic:2012jj}. The template overlap method 
is designed for use in boosted top identification as well as boosted Higgs 
identification.  The method is similar to that of shower deconstruction, 
in that it attempts to quantify how well a given jet matches a certain 
expectation such as a boosted top quark or boosted Higgs decay. However, 
this method uses only final state configurations, whereas the shower 
deconstruction method takes into account the showering histories. A catalog 
of templates is formed by analyzing signal events. Once this is in place, 
individual jets can be analyzed by evaluating an overlap function which 
evaluates how well the current jet matches the templates from the signal 
process of interest.  For example, a template for hadronic boosted top 
quark decays would consist of three energy deposits within the jet. In 
studies with high-$p_T$ jets, the rejection factor of QCD jets compared 
to jets from boosted top quark decays is of the order 10$^2$. One additional 
feature of this template overlap method is the automatic inclusion of 
additional parton radiation into the template catalog, such as for Higgs 
decays to bottom quark pairs, where there is commonly an additional gluon 
radiated, resulting in 3 energy deposits instead of the 2 from the $b$ quarks.

Finally, the Q-jets~\cite{Ellis:2012sn} scheme could be used for top-tagging.
This is a method to remove dependence of analysis results on the choice of 
clustering algorithm used to reconstruct jets.  For example, one could use 
either the Cambridge-Aachen algorithm or the kT algorithm to cluster jets, 
and may obtain significantly different results in the jet masses.  The 
Q-jets algorithm attempts to use all possible ``trees'' to cluster 
constituents, rather than using the single tree provided by the specific 
clustering algorithm used.  In this way, each jet now has a distribution 
of possible masses instead of a single jet. This provides additional 
information which can enhance signal discrimination. For example, the 
variance of the jet mass between individual clustering trees can be examined, 
rather than relying on just a single value. The statistical stability is 
also enhanced when using the Q-jets algorithm.

\subsection{Searches with boosted top quarks}

The first area where new tools developed specifically for the selection and 
reconstruction of boosted top quarks have shown their value is in searches for 
massive new states decaying to top quark pairs. The first application 
of techniques specifically aimed at boosted top decays was the CMS 
\ttbar{} resonance search in the all-hadronic channel~\cite{Chatrchyan:2012ku}.
The evolution of the 
mass reach~\footnote{The sensitivity to massive particles is expressed in terms
of the observed 95 CL lower limit on the mass of a leptophobic topcolor \Zprime{} 
boson. The motivation of this particular model may not have survived recent 
advances in particle physics, but to monitor the sensitivity
of searches it is still the best benchmark on the market.} 
of \ttbar{} resonance searches in the more sensitive 
``lepton+jets'' channel is shown in 
Fig.~\ref{fig:summary_searches}. 
By the start of the LHC program the Tevatron experiments had excluded a \Zprime{} boson
mass lower than 700~\gev{}~\cite{Aaltonen:2007ag,Abazov:2008ny}. In the course of 2011 and 2012 
the limit was extended to 800~\gev{} by a D0 search on nearly 5~\ifb{}~\cite{Abazov:2011gv} 
and to approximately 900~\gev{} by a CDF analysis of
the complete Tevatron data set~\cite{Aaltonen:2012af}.
An ATLAS search on 2.4~\ifb{} of 7~\tev{} LHC data~\cite{Aad:2012wm} collected 
in 2011 reached a
similar precision. All these analyses followed the conventional, {\it resolved}
approach that is based on the assumption that the six fermions from the decay of
the top quark pair
($t \rightarrow W^+ b \rightarrow l^+ \nu_l b$ and the charge conjugate process)
can be resolved individually.

In some cases ATLAS and CMS analyses specifically designed for boosted top quarks~\cite{Aad:2012dpa,Chatrchyan:2012cx} scrutinized the same data set that had been used by the {\it resolved} approach.
A direct comparison of these results demonstrates that the novel approach has considerably better sensitivity for massive 
states~\cite{Aad:2012dpa}. The final analyses on 2011 data~\cite{Aad:2013nca,Chatrchyan:2012cx} combine {\it resolved}
and {\it boosted} methods to attain good 
sensitivity over the complete mass spectrum. The excluded mass range is pushed 
up to 1.74~\tev{}.

Searches in the ``lepton+jets'' channel are complemented by analyses of the fully hadronic
($t\bar{t} \rightarrow 6$ jets) and di-lepton ($t\bar{t} \rightarrow b \bar{b} l^+ \nu_l l'^- \bar{\nu}_l'$) decay chains. 
Only one fully hadronic \ttbar{} resonance search was performed at the Tevatron~\cite{Aaltonen:2011vi}.
At the LHC, with a daunting multi-jet background, these searches are even more challenging. 
The advent of new algorithms has, however, greatly boosted their potential. The mass reach of the
CMS~\cite{Chatrchyan:2012ku} and ATLAS search~\cite{Aad:2012raa} are compared to that of the
``lepton+jets'' searches in Table~\ref{tab:limits}. 

The prospects for progress are good. Preliminary results on the 2012 data 
set~\cite{ATLAS-CONF-2013-052,CMS-PAS-B2G-12-005,CMS-PAS-B2G-12-006} 
have significantly extended previous limits.

\subsection{Jet substructure performance and searches}

The results in the previous Section demonstrate the proof-of-principle: 
the addition of jet substructure to the experimentalists' tool-box 
boosts the sensitivity of searches for new physics at the LHC. 
It is clear, however, that these tools are
still in their infancy. In all searches discussed in the previous Section 
large systematic uncertainties are assigned to the large-R jets. 
It is natural to suspect that further progress could be made with better 
(and, especially, better understood) tools. 

To quantify the impact of the jet-related systematics on the sensitivity we
have evaluated expected limits on the narrow \Zprime{} boson with all sources
of systematic uncertainty, except one (so-called $N-1$ limits) in several
iterations of the ATLAS searches in the lepton+jets final state. The
uncertainties associated with the large-R jet that captures the 
hadronic top decay are always the dominant source of uncertainty. 
Their impact is considerably larger than that of systematics 
associated with the narrow jets, even at relatively low resonance mass.
The limits
over a large mass range (1-2~\tev{}) would improve by approximately 5-10\% if 
only the uncertainty on the scale and resolution of mass and energy of
anti-$k_t$ jets with $R=1$ is removed. 

If we apply an ad hoc scale factor
of two to this uncertainty (representing a failure to bring these
uncertainties under control) we find that the sensitivity is further degraded. 
A significant reduction of large-R 
jet uncertainties, on the other hand,
brings the $N-1$ limits with no jet-related systematics and the 
limits with reduced large-R jet systematics to within
2\%. 

CMS has not published the $N-1$ results for their searches, but 
qualitatively the same picture emerges. In the fully
hadronic searches the jet-related uncertainties have the largest
impact on the limits.

We conclude that further progress undertanding jet substructure 
still has substantial
potential to increase our sensitivity to massive new states decaying 
to top quarks.

\subsection{Further applications}

The selection for boosted top quarks, in the lepton+jets and fully hadronic
channels, have proven their value in \ttbar{} resonance searches, but
are more generally applicable. 

The obvious direction to extend the 
range of applications is to other searches with boosted top quarks.
The $W' \rightarrow tb$ that are currently performed in the channel
where the top quark decays to a charged lepton, neutrino and b-jet.
We expect, however, that, ultimately the highest mass reach should
be obtained in the hadronic decay (with a factor two large branching
ratio if $\tau$-leptons are not considered).

We expect differential cross-section measurements for \ttbar{} to benefit
from these techniques at large transverse momentum and invariant mass
of the \ttbar{} pair. Apart from the better selection efficiency in
algorithms designed for this kinematic regime, the 
better truth-to-reconstructed mapping of $p_T$ and $m_{\ttbar}$ 
is expected to be an important advantage. We are looking forward to 
such measurements from the ATLAS and CMS experiments.
Also analyses that rely
strongly on the reconstruction of the top quark direction, such as
the charge asymmetry measurement, should benefit.

Finally, several authors~\cite{Plehn:2009rk} have commented on the potential
of events with mildly boosted top quarks for the observation of $t\bar{t}H$
and a measurement of the  production rate.

\subsection{Summary}

Over the last five years, many ideas have been proposed to cope with the 
challenge of boosted top quark reconstruction. Since then, these ideas
have been implemented by the experiments and put to the test, primarily
in searches for massive new states decaying to \ttbar{} pairs. The overview
we presented in Fig.~\ref{fig:summary_searches} and Table~\ref{tab:limits}
is a testimony to the increase of sensitivity for such states fuelled
by the performance of the LHC.
Such progress would not have been possible if novel techniques for the 
study of boosted top quarks had not been developed.
We expect the selection developed for the lepton+jets and fully hadronic
to find further applications in searches and measurements.


\section{Summary \& Conclusions}
\label{sec:conclusions}
This report of the BOOST2012 workshop provides answers to a number of 
important questions concerning the use of jet substructure for the 
study of boosted object production at the LHC.

We evaluated the current limitations in the description of jet substructure,
both at the analytical level and in Monte Carlo generators. 
Impressive progress is being made for the former and we expect a meaningful 
comparison to LHC 
data to be a reality soon. Two approaches - perturbative QCD and
Soft Collinear Effective Theory - to a first-principle resummation of
the jet invariant mass are producing mature results. Measurements
of the jet mass in Z+jet events are proposed, both inclusively and
exclusively in the number of jets. We hope that in
the not-too-distant future these calculations can enhance our
understanding of the internal structure in jets.

Monte Carlo predictions remain crucial to searches and measurements
employing jet substructure. We have compared the predictions of
several mainstream generators for a number of substructure observables a
and for several signal and background topologies.
While jet mass is still poorly described
by several generators, several ways of introducing the inherent
uncertainties become evident. Jet grooming reduces the spread among
Monte Carlo models, as do several alternative jet substructure 
observables.

We also studied potential experimental limitations that could check further
progress, in particular the impact of the large number of simultaneous
proton-proton interactions. We find that, even if the substructure of
large-radius jets is quite sensitive to pile-up, a combination of 
a state-of-the-art correction technique and jet grooming can effectively
restore the jet mass scale and strongly mitigate the impact on the 
jet mass resolution.

Finally, we reviewed top-tagging techniques deployed in the LHC experiments
and assessed their impact on the sensitivity to new physics. A series
of \ttbar{} resonance searches performed by ATLAS and CMS provide
clear proof of the power of techniques specifically designed for 
boosted top quarks. Through an evaluation of the impact of all
sources of systematic uncertainties, we show that further progress
can still be made with an enhanced understanding of jet substructure.
We expect to see these techniques applied in further searches involving
boosted top quarks and in measurements of the boosted top production
rate.

\section*{Acknowledgements}

We thank the Spanish Center for Particle Physics, Astroparticle and 
Nuclear Physics (CPAN), 
the regional government (Generalitat Valenciana), Heidelberg University 
and IFIC (U. Valencia/CSIC) for their generous support of the BOOST2012
conference. 

We furthermore thank the Fundaci\'on Cultural Bancaja for putting 
at our disposal 
the ``Centro Cultural'' that hosted the workshop, the IT 
teams at IFIC, UW and LPTHE for the facilities offered to host
the BOOST samples, TotNou for the organization of the workshop, 
Isidoro Garc\'ia of CPAN for organizing the outreach event 
and coordinating the contacts with the media and 
Pilar Ordaz for the design of the poster and logotype.


\bibliographystyle{JHEP}       
\bibliography{boost2012_report}   

%
%

\end{document}